\documentclass[journal]{IEEEtran}
\usepackage[LGR,T1]{fontenc}
\usepackage[latin9]{inputenc}
\usepackage{varwidth}
\usepackage{relsize}
\usepackage{amsmath}
\usepackage{amssymb}
\usepackage{graphicx}
\usepackage{enumitem}
\usepackage{cite}
\usepackage[hidelinks]{hyperref}

\makeatletter

\providecommand{\tabularnewline}{\\}
\newenvironment{cellvarwidth}[1][t]
    {\begin{varwidth}[#1]{\linewidth}}
    {\@finalstrut\@arstrutbox\end{varwidth}}

\usepackage{array}
\usepackage{multirow}
\usepackage{xcolor}

\makeatother

\begin{document}
\title{RF sensing with dense IoT network graphs:\\ An EM-informed analysis}

\author{\IEEEauthorblockN{Federica Fieramosca, Vittorio Rampa,
Michele D'Amico, Stefano Savazzi} 

 \thanks{Funded by the European Union (EU). Views and
opinions expressed are however those of the author(s) only and do not necessarily reflect those of the EU or European Innovation Council and SMEs Executive Agency (EISMEA). Neither the European Union nor the granting authority can be held responsible for them. Grant Agreement (GA) No: 101099491 (Holden project). \\F. Fieramosca and M. D'Amico are with the Dipartimento di Elettronica, Informazione e Bioingegneria (DEIB), Politecnico di Milano, 20133 Milan, Italy (e-mail: federica.fieramosca@polimi.it, michele.damico@polimi.it). \\
V. Rampa and S. Savazzi are with
the Consiglio Nazionale delle Ricerche, Institute of Electronics, Information Engineering and Telecommunications (IEIIT), 20133 Milan, Italy (e-mail: vittorio.rampa@cnr.it, stefano.savazzi@cnr.it).}}


\maketitle
\begin{abstract}
Radio Frequency (RF) sensing 
is attracting interest in research,
standardization, and industry, especially for its potential in Internet of Things (IoT) applications. By leveraging the properties of the ElectroMagnetic (EM) waves used in wireless networks, RF sensing captures environmental information such as the presence and movement of people
and objects, enabling passive localization and vision applications.
This paper investigates the theoretical bounds on accuracy and resolution for RF sensing systems within dense networks. It employs an EM model to predict the effects of body blockage in various scenarios. To detect human movements, the paper proposes a deep graph neural network, trained on Received Signal Strength (RSS) samples generated from the EM model. These samples are structured as dense graphs, with nodes representing antennas and edges as radio links. Focusing on the problem of identifying the number of human subjects co-present in a monitored area over time, the paper analyzes the theoretical limits on the number of distinguishable subjects, exploring how these limits depend on factors such as the number of radio links, the size of the monitored area and the subjects physical dimensions. These bounds enable the prediction of the system performance during network pre-deployment stages. The paper also presents the results of an indoor case study, which demonstrate the effectiveness of the approach and confirm the model's predictive potential in the network design stages.
\end{abstract}


\IEEEpeerreviewmaketitle{}

\section{Introduction}

\IEEEPARstart{RF}{} sensing refers to a set of opportunistic techniques designed to detect, locate, and track people within areas covered by ambient Radio Frequency (RF) signals~\cite{wilson-2010}. Aligned with the goals of \emph{transformative computing}, i.e., reusing communication networks for concurrent sensing task~\cite{savazzi-1}, and the \emph{communication-while-sensing} framework \cite{isac_survey}, RF sensing systems transform each wireless network node into a \emph{virtual sensor}, enabling both communication and environmental perception capabilities. Typically operating in the unlicensed $2.4\div6$~GHz Industrial, Scientific, and Medical (ISM) bands for indoor applications, these systems leverage the cm-scale wavelengths of RF signals to capture body-induced alterations in multipath scenarios.

\begin{figure}
\centering\includegraphics[width=1\columnwidth]{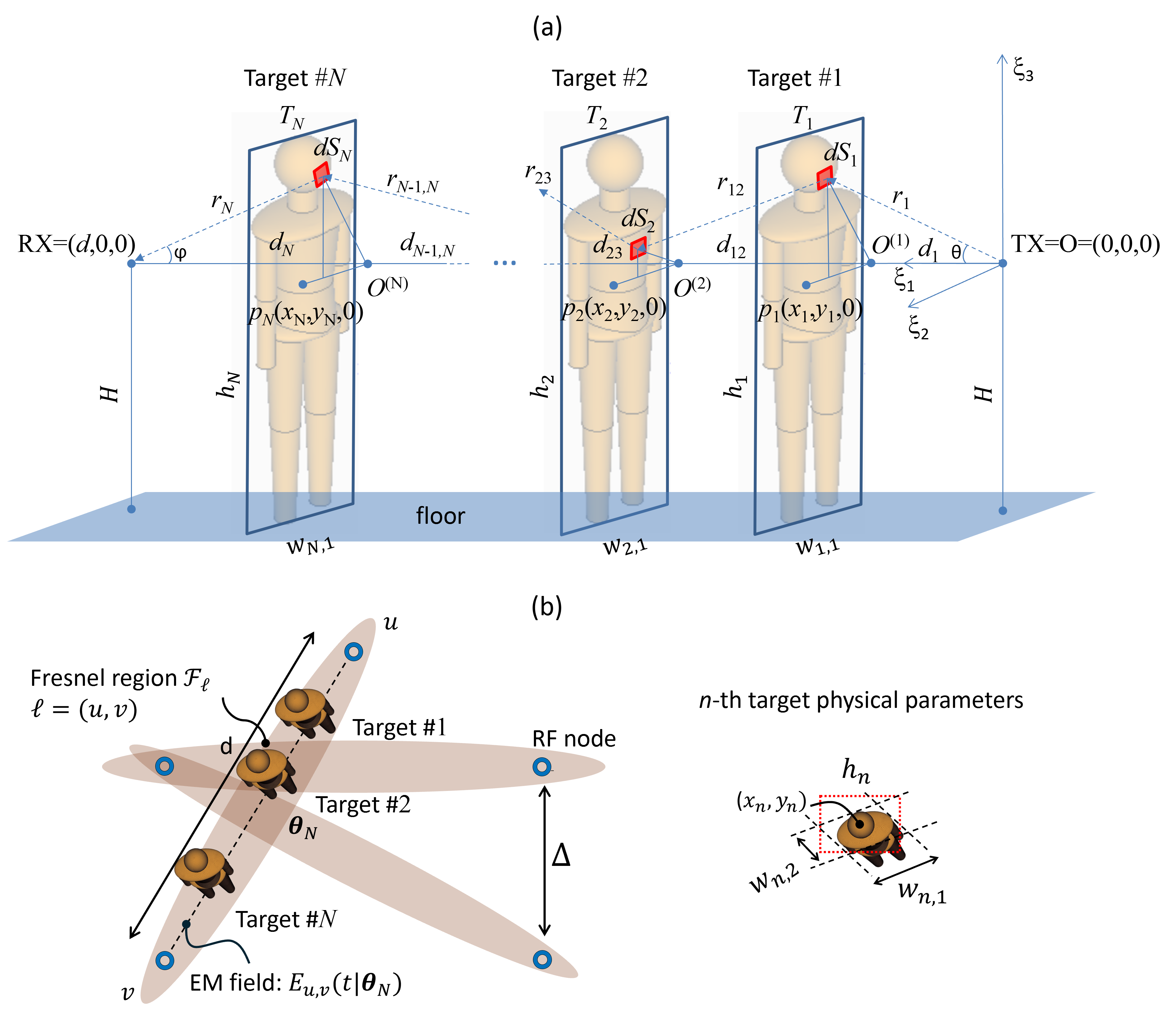} \caption{\label{antodiffr} RF sensing scenarios: (a) Single-link multi-target
scenario consisting of a single link of length $d$, horizontally placed at
a height $H$ above the floor. The scenario includes $N$ different targets
located at $p_{1}(x_{1},y_{1},0)$, $p_{2}(x_{2},y_{2},0)$,
\dots , $p_{N}(x_{N},y_{N},0)$, respectively. (b) A generalization of case (a), where multiple links are considered in a generic setup. The targets are now positioned at random locations within the
area, each characterized by parameters such as height $h_n$, orientation
$\varphi_n$, anteroposterior and lateral body dimensions $w_{n,1}$ and $w_{n,2}$, respectively. The goal is to predict the number of subjects co-present from the analysis of the received EM field $E_{u,v}$.}\label{new_intro}
\end{figure}

Unlike active RF sensing techniques, which require people to carry specific RF devices, we focus here to passive RF techniques, also known as device-free, to sense, localize, and track people in areas covered by ambient RF signals due to already available communication devices. These techniques~\cite{shit-2019}, such as Bayesian filtering, tomographic imaging,
and RF holography~\cite{kat,fall,holog}, require a detailed understanding
of body blockage effects and their EM wave interactions, which prevent the adoption of classical supervised Machine Learning
(ML) methods. Moreover, applying RF sensing at scale demands solving complex inverse problems on high dimensional graph structures, a fundamentally ill-posed estimation challenge~\cite{geom}.

The influence of body movements on RF signals can be predicted
through considerations rooted in EM wave propagation theory. Recently,
numerous physical and statistical models have been introduced, exploiting
various methods such as ray tracing~\cite{phyindoor,rayback}, moving
point scattering~\cite{scatt}, Knife-Edge Diffraction (KED)~\cite{qi-2017}, Geometrical Theory of Diffraction (GTD)~\cite{james-2007,qi-2017}, Uniform Theory of Diffraction (UTD)~\cite{peter-2012,plouhinec-2023}, and Scalar Diffraction Theory~\cite{rampa-2017,rampa-2022a,j-fieramosca-2023}.
In addition, to strongly reduce the simulation time, generative neural
networks have recently emerged as promising tools for fast
field computation and to solve inverse imaging problems~\cite{generative_mag},
when properly trained with off-line simulation results~\cite{generation}.

\subsection{RF sensing scenario: infrastructural snapshot}

\label{subsec:RF_sensing _cenario}

This paper considers a dense deployment of wireless communication
radio devices (i.e., nodes) that collect RF measurements to reconstruct an \textit{infrastructural snapshot}~\cite{moral}
of the environment. As illustrated in Fig.~\ref{new_intro}, the goal is
to obtain a precise prediction of the number of subjects (i.e., the targets) moving in the considered space in each time instant $t$, 
by leveraging the same wireless communication services they provide. This form of environmental imaging enables real-time monitoring of the state and usage of a given environment, allowing the identification of emerging or recurring body-movement patterns, with relevant applications in smart spaces, assisted living, and safety~\cite{avola,determe}. The problem of people counting using passive RF sensing has been approached from various perspectives in the literature. For example, \cite{cianca} introduced a modular approach for queue counting, achieving more than $90$\% accuracy. Fine-grained crowd and speed estimation has also been studied in~\cite{hao} through passive WiFi sensing. More recently, most crowd analysis works rely on data-driven strategies for training machine learning models~\cite{access_rev}. However, none of these studies have explicitly addressed the crucial issues of quantifying resolvable bounds for target discrimination or designing the RF sensing network prior to deployment with the aid of EM body models. These aspects are instead crucial to establish the theoretical limits of passive RF sensing and to guide the practical design of sensing infrastructures.

We consider a multi-target scenario where $N>1$ targets move randomly within the monitored area $\mathcal{X}$. Targets may freely enter or exit
the environment, so the number of active targets $N(t)$ is time-varying as in~\cite{hao}. The physical dimensions of the targets and the overall crowd size significantly affect body diffraction phenomena and thus the sensing and imaging performances. Accordingly, each target is characterized
by distinct physical features, collected as: $\mathbf{\boldsymbol{\theta}}_{N}=[\theta_{1},...,\theta_{n},...,\theta_{N}]$,
with $\mathbf{\theta}_{n}:=\left\{ \mathbf{p}_{n},\varphi_{n},h_{n},w_{n,1},w_{n,2}\right\} $,
where $\mathbf{p}_{n}=(x_{n},y_{n},z_{n})$ denotes the target location, $\varphi_{n}$
its orientation, $h_{n}$ its height, and $w_{n}$ its effective traversal size. The latter is bounded in a 3D cylindrical representation~\cite{generation} as $w_{n,2}\le w_{n}\le w_{n,1}$,
with $w_{n,1}$ and $w_{n,2}$ denoting the anteroposterior and lateral body dimensions, respectively. As shown in Fig.~\ref{new_intro}, $z_{n}$ simplifies to $z_{n}=0$ due to the chosen coordinate system and origin, and will therefore be omitted in the following so that $\mathbf{p}_{n}=(x_{n},y_{n})$.

\subsection{Contributions}
\label{subsec:contributions}

RF signals are modeled using graph structures, where nodes correspond to radio antennas and edges denote active communication links. 
Each node is associated with high-dimensional \textit{node features}, derived from the Channel State Information (CSI) received from its neighbors at time $t$. In their simplest form, CSI features can be expressed through Received Signal Strength (RSS) values, which carry information
on the presence of concurrently moving subjects within the node neighborhood. Each graph thus provides a \textit{snapshot}
of the environment at a specific time instant $t$. The objective is to estimate the number of moving targets, $N(t)$, from previously unseen graph observations at any time $t$.

The graph representation of RF signals naturally lends itself to processing with Graph Neural Network (GNN), which are specifically designed for graph-structured data, relational and topological data analysis. Consequently, the crowd sensing (target counting) is framed as a graph classification problem~\cite{duev,lei}.
The presence of targets alters the RF field, leading to populations of graphs and node features that encode information about
the target location, size and movement. Building on multi-body EM modeling,
this paper establishes practical limits on the maximum number of targets
that can be reliably detected and distinguished. These limits depend on the
RF graph structure, the number of links, the
physical characteristics of the subjects and environment, and the signal wavelength. The theoretical bounds are derived via stochastic
geometry and validated through a flexible classification framework based on a Deep Graph Convolutional Neural Network (DGCNN), which applies convolutional operations over the graph structure. The DGCNN model is trained on synthetic data from EM body model samples. In summary,
the key contributions of this paper are:
\begin{itemize}
\item \emph{Multi-body EM model}: the Composite Multi-body Additive Model (C-MAM) is proposed, a simplified model that captures the cumulative RF effects of multiple subjects in dense sensing networks, explicitly accounting for variable RF links/edges, antennas/nodes
and coverage. Unlike existing models ~\cite{eibert,smith,pravadelli,depatla,yang}, C-MAM is built on classical diffraction theory and captures the combined RF effects of multiple targets in dense sensing networks, explicitly incorporating EM interference caused by co-located subjects. 
\item \emph{Resolvable target bounds}: based on the above model, the paper derives practical limits on the number of
targets that can be reliably distinguished in dense graph settings, where the number
of edges approaches the network's connectivity limits. These bounds are linked to the geometric configuration of RF links, with target positions modeled as a stochastic
process. They serve as a tool for pre-deployment performance assessment across diverse propagation scenarios and body characteristics. For multi-link single-target pre-deployment strategies, the interested reader can refer to~\cite{kianoush-2016} as well.
\item \emph{DGCNN-based classification tool}: a classification framework using a DGCNN architecture is designed, incorporating prior EM knowledge, thereby reducing the need for supervised learning and mitigating overfitting impairments. 
The proposed tool is validated both in simulation and through an indoor experimental campaign. A comparative performance analysis is carried out among models trained on: (i) experimental data, (ii) synthetic data generated using C-MAM, (iii) synthetic data obtained from existing state-of-the-art multi-body physical models, and (iv) the corresponding resolvable target bounds.

\end{itemize}
The paper is organized as follows. Sect.~\ref{sec:syst} introduces the system model and the graph-based representation. Sect.~\ref{sec:em_model}
discusses various multi-body EM models and their role in dense RF
networks. Sect.~\ref{sec:limits} derives practical resolvability bounds, which are then validated via a DGCNN-based classification tool in Sect.~\ref{sec:Graph-Neural-Network}. Sect.~\ref{sec:Numerical-results-and} and \ref{casestudy}
present results from simulated environments and an indoor case study at $2.4$~GHz, respectively, thus demonstrating the predictive performance of the proposed model. Concluding remarks and open research directions are drawn in Sect.~\ref{sec:Conclusions}.

\section{System model \label{sec:syst}}

The wireless network system considered in this work is illustrated in Fig.~\ref{new_intro}.
The network consists of an arbitrary set $\mathcal{V}$ of RF nodes (or antennas), strategically positioned around the monitored area $\mathcal{X}$
and a corresponding set of active radio links $\mathcal{L}$, where
$\left|\mathcal{L}\right|=L\leq\left|\mathcal{V}\right|(\left|\mathcal{V}\right|-1)$. 
These links collectively ensure the coverage of the monitored area. The nodes and links define a
graph structure $\mathcal{G}=(\mathcal{V},\mathcal{L})$, where $\mathcal{V}$
and $\mathcal{L}$ represent the set of vertices and edges of size $\left|\mathcal{V}\right|$ and $\left|\mathcal{L}\right|$, respectively. Each edge
corresponds to an active physical radio link $\ell=(u,v)\in\mathcal{L}$
between antennas $u\in\mathcal{V}$ and $v\in\mathcal{V}$, with
$u\ne v$. All active links maintain sufficiently strong signal
levels to support reliable communication, even in the presence of obstructions
caused by the targets.

Each node $u$ is located at coordinates $p_{u}=\left(x_{u},y_{u},z_{u}\right)$
and operates either as a transmitter (TX) or receiver (RX), depending on the network's
connectivity design. The set of active neighboring nodes for a given node $u\in\mathcal{V}$ is denoted by $\mathcal{N}(u)=\left\{ v\in\mathcal{V}:(u,v)\in\mathcal{L}\right\}$ with size $\left|\mathcal{N}(u)\right|$.
A time-division access scheme governs the network communications, allowing each antenna $u$ to measure, at time $t$, the EM field
of RF beacon signals transmitted by its neighbors according to: 
\begin{equation}
\mathbf{E}_{u,t}(\mathbf{\boldsymbol{\theta}}_{N}):=\left\{ E_{u,v}(t|\boldsymbol{\theta}_{N})\right\} _{v\in\mathcal{N}(u)}\in\mathbb{R}^{\left|\mathcal{N}(u)\right|},\label{eq:features}
\end{equation}
where measurements may be affected by the presence of $N$ targets characterized by parameters $\boldsymbol{\theta}_{N}$, previously introduced in Sect.~\ref{subsec:RF_sensing _cenario}. The resulting vector $\mathbf{E}_{u,t}$
contains the node \emph{features} for \emph{ $u$} and serves as input
for processing. 
By aggregating measurements across all nodes, the full set at time $t$ forms a matrix of size $\left|\mathcal{V}\right|\times\mathrm{max}_{u\in\mathcal{V}}\left|\mathcal{N}(u)\right|$, defined as: 
\begin{equation}
\mathbf{E}_{t}(\mathbf{\boldsymbol{\theta}}_{N})=\left\{ \mathbf{E}_{u,t}(\mathbf{\boldsymbol{\theta}}_{N})\right\} _{u\in\mathcal{V}}\label{eq: node features}
\end{equation}
Modeling of the EM field samples $E_{u,v}(t|\mathbf{\boldsymbol{\theta}}_{N})$ in terms of body-induced field attenuations ($A_{u,v}$) for each link is discussed in the next section.
For simplicity, but without any loss of generality, in the following sections we assume that all nodes are placed on a horizontal plane at height $H$ above the floor (see also Fig.~\ref{new_intro}).

\section{EM modeling of multiple human bodies blockage effects\label{sec:em_model}}

This section describes the single-target diffraction model,
and its extension to a multiple obstacle scenario. Two multi-body approaches are verified and compared, namely the Multi-body Additive Model
(MAM) and the Composite Multi-body Additive Model (C-MAM). All models
are defined for a single-link scenario in Sects.~\ref{subsec:sodm},
and~\ref{subsec:mam} and~\ref{subsec:smam}, and then extended for a generic multi-link layout.

\subsection{Single-Obstacle Diffraction Model}
\label{subsec:single}

\label{subsec:sodm}

The single-obstacle model, already presented in~\cite{rampa-2017}, is based on the scalar diffraction theory and calculates the attenuation due to a single target by considering its effect on the transmitted field. 
The scalar diffraction theory is based on Huygens' principle~\cite{baker-2003}, which states that each point of a wavefront (i.e. a surface where the fields have the same phase) acts as a (virtual) source of a secondary spherical wavelet, and these wavelets combine to produce a new wavefront in the direction of propagation. In other words, waves propagate forward as if every point along the front is re-emitting the wave. These virtual sources are called \textit{Huygens' sources}.
The presence of a single target (i.e., $n=1$),
modeled as a 2D absorbing sheet $T_n$, introduces attenuation in the
signal transmitted from the node $u\in\mathcal{V}$ and received by
the node $v\in\mathcal{V}$ by blocking a fraction of the Huygens' sources. 
The sources that contribute most to the received power at RX (and hence to attenuation if blocked) are those included in the first Fresnel's region, a prolate ellipsoidal region of space around the radio link, with transmitter (TX) and receiver (RX) located at its foci, therefore named also first Fresnel's ellipsoid ~\cite{bookfresnel}. Using the same notations shown in Fig.~\ref{new_intro}, the received electric field $E_{u,v}(t|\theta_{n})$
for the link $\ell=(u,v)\in\mathcal{L}$ is expressed, w.r.t. the received
electric field $E_{u,v}(\emptyset )$ in \textit{free-space} (i.e., with no target in the link area), as:

\begin{equation}
\frac{E_{u,v}(t|\theta_{n})}{E_{u,v}(\emptyset )}=1-j\frac{d}{\lambda}\int_{T_{n}}\frac{1}{r_{1}r_{2}}\exp\left\{ -j\frac{2\pi}{\lambda}(r_{1}+r_{2}-d)\right\} dT,
\end{equation}
where $\theta_{n}$ corresponds to the features of the only target $T_n$
present in the link area, $d$ is the straight distance between $u$
and $v$, $\lambda$ the wavelength of the RF signal, $r_{1}$ and
$r_{2}$ the distances from an elementary area of size $dT$ on the
obstacle $T_n$ to the transmitting and receiving nodes, respectively.

The body-induced attenuation $A_{u,v}$ w.r.t. the free-space scenario caused by the single obstacle $T_n$ is quantified in decibel (dB) as $A_{u,v}=A_{T_n}$ where:

\begin{equation}
A_{T_n}=-10\log_{10}\left|\frac{E_{u,v}(t|\theta_{n})}{E_{u,v}(\emptyset )}\right|^{2}.
\label{eq:AT-db}
\end{equation}

\noindent This model has been evaluated in several field measurement trials and has shown to provide a good foundation~\cite{rampa-2017,j-fieramosca-2023,kianoush-2016} for understanding how a single obstacle modifies the EM field and influences the RSS. Please note that, in the following sections, unless mentioned, the term attenuation refers to the attenuation w.r.t. the free-space scenario.

\subsection{Multi-body Additive Model (MAM)}

\label{subsec:mam}

The MAM model is a straightforward extension of the single target model of Sect.~\ref{subsec:single} to multiple targets (i.e., $n>1$) within a single link. The models assumes that the total attenuation is the sum of the individual contributions of all targets. This approach has been exploited in~\cite{nicoli-2016,rampa-2021} for localization of multiple targets.  
Other works have employed similar additive hypothesis to predict the attenuation of multiple bodies ~\cite{wilson-2010,bocca-2014}. In MAM frameworks, each target $T_{n}$ contributes to the total attenuation $A_{u,v}$ as follows:

\begin{equation}
A_{u,v}=\sum_{n}A_{T_{n}},
\end{equation}
where $A_{T_{n}}$ represents the attenuation caused by the $n$-th single target that is present in the area of the link $(u,v)$ according to (\ref{eq:AT-db}). This approach, referred to as Multi-body Additive Model (MAM), captures
the cumulative effects of multiple targets within the link. 

Considering (\ref{eq:features})
and (\ref{eq: node features}), the attenuation samples measured by
antenna $u$ at snapshot $t$ are organized as the feature vector:
\begin{equation}
\mathbf{A}_{u,t}:=\left\{ A_{u,v}\right\} _{v\in\mathcal{N}(u)}.\label{eq:features-att}
\end{equation}
Similarly, considering all the nodes, the matrix: $\mathbf{A}_{t}=\left\{ \mathbf{A}_{u,t}\right\} _{u\in\mathcal{V}}$ collects all the network samples at time $t$.

\subsection{Composite Multi-body Additive Model}

\label{subsec:smam}

In what follows, we propose a composite model, referred to as Composite MAM (C-MAM), which extends classical multi-body diffraction by integrating additional phenomena relevant to multi-subject scenarios. The MAM model of Sect.~\ref{subsec:mam} is further refined
to account for two key considerations: 
\begin{enumerate}
\item the attenuation due to a target is assumed to be negligible if the target lies
outside the first Fresnel's ellipsoid of the link; 
\item if a target completely obstructs the first Fresnel's ellipsoid of a radio link, 
then additional contributions from other obstacles, namely outside the first Fresnel's ellipsoid, are negligible and the total attenuation is equal to the attenuation
caused by the obstructing target. 
\end{enumerate}
The above assumptions will be verified by the experiments in Sects.~\ref{sec:Numerical-results-and} and~\ref{casestudy}. Based on the first consideration, the attenuation
$A_{T_{n}}$ for a target $T_{n}$ in a given link is modified as~\cite{geom}: 
\begin{equation}
A_{T_{n}}=\begin{cases}
-10\log_{10}\left(\left|\frac{E_{u,v}(t,\theta_{n})}{E_{u,v}(\emptyset )}\right|^{2}\right), & \text{if }T_{n}\in\mathcal{F}_{\ell},\\
0, & \text{otherwise.}
\end{cases}
\label{eq:C-MAM-AT}
\end{equation}
where $\mathcal{F}_{\ell}$ indicates the Fresnel's region of the link $\ell:=(u,v)\in\mathcal{L}$.

Moreover, according to the second assumption, if the target intersects
the main ray between $u$ and $v$, the total attenuation associated
to the link is given by:

\begin{equation}
A_{u,v}=\begin{cases}
\max\left\{ A_{T_{n}}\right\} , & \text{if }\exists\,T_{n}:T_{n}\in\mathcal{F}_{\ell}\\
0, & \text{otherwise,}
\end{cases}\label{eq:7}
\end{equation}
where $A_{T_{n}}$ represents the attenuation (\ref{eq:AT-db}) caused by the obstructing
single target $T_{n}$. Eq. (\ref{eq:7}) establishes that the attenuation of the
link between $u$ and $v$ is dominated by the target $T_{n}$ whose
position and dimensions block the link first Fresnel's ellipsoid the most. 
In other words, this condition ensures that when a target substantially blocks the first Fresnel's ellipsoid, other targets located in the same area do not introduce any additional attenuation, as is logical, since they are in the \textit{shadow} of the dominant target. This condition prevents unrealistically high overall attenuation values from being estimated, which would happen if the contributions of the individual targets were simply added together.

It is worth noticing that, the feature vector $\mathbf{A}_{u,t}$ is still obtained as indicated in (\ref{eq:features-att}) using now the C-MAM model described by (\ref{eq:C-MAM-AT}) and (\ref{eq:7}). Likewise, $\mathbf{A}_{t}=\left\{ \mathbf{A}_{u,t}\right\} _{u\in\mathcal{V}}$ collects all the network samples at time $t$.

\section{Limits on resolvable subjects in dense graphs}

\label{sec:limits}

This section investigates the practical limits on the number of resolvable
targets based on the EM diffraction analysis previously discussed.
We consider a network 
modeled as a
dense graph with $L$ edges as described in Sect.~\ref{sec:syst}. The proposed limit can serve as an upper bound
to quantify the number of resolvable targets as well as the effects
of the antenna (graph) density, namely the number of active radio links
$L$, the size of the monitored area $\mathcal{X}$ and the frequency,
that determines the size of the Fresnel's region $\mathcal{F}_{\ell}$ of a particular link $\ell$.
As shown in the analysis of the next section, the limit is valid under dense graph structures for which the number of edges approaches
the maximum, namely $L\simeq\left|\mathcal{V}\right|(\left|\mathcal{V}\right|-1)$.

We assume $N$ targets positioned on a 2D plane. These targets are characterized
by different physical characteristics: $\mathbf{\boldsymbol{\theta}}_{N}=[\theta_{n}]_{n=1}^{N}$
and relative locations $\mathbf{p}_{n}=(x_{n},y_{n})$. We define
the indicator function as $\mathbf{1}_{A}=1$ iff condition $A$ is
true and $\mathbf{1}_{A}=0$ otherwise. For example, to indicate if
a target $T_n$ is inside the Fresnel's region $\mathcal{F}_{\ell}$, we use the following notation $\mathbf{1}_{\mathbf{p}_{n}\in\mathcal{F}_{\ell}}$, defined as:
\begin{equation}
\mathbf{1}_{\mathbf{p}_{n}\in\mathcal{F}_{\ell}}=\begin{cases}
1 & \text{if }\mathbf{p}_{n}\in\mathcal{F}_{\ell}\text{ ($T_n$ inside Fresnel's reg. of \ensuremath{\ell})}\\
0 & \text{if }\mathbf{p}_{n}\notin\mathcal{F}_{\ell}\text{ ($T_n$ outside Fresnel's reg. of \ensuremath{\ell}).}
\end{cases}
\end{equation}
\noindent In particular, we consider a target $\mathbf{p}_{n}$ to belong to a Fresnel's region $\mathcal{F}_{\ell}$ if at least $50\%$ of its footprint\footnote{The target area footprint can be approximated by an elliptical area, derived from the lateral and anteroposterior body sizes: $\pi\cdot w_{n,1}/2\cdot w_{n,2}/2$. } lies within that region. The $50\%$ overlap threshold was chosen empirically as a reasonable trade-off: it includes targets that are substantially enclosed by the Fresnel's region, while avoiding noisy or ambiguous assignments. Lower thresholds led to unstable classifications, whereas stricter ones excluded partially relevant targets. We also define the set $Q_{n}$ as: 
\begin{equation}
Q_{n}:=\{\ell\in \mathcal{L}:\mathbf{p}_{n}\in \mathcal{F}_{\ell}\}.
\end{equation}
This represents the subset of the links for which the $n$-th target relative
position $\mathbf{p}_{n}$ falls within the corresponding Fresnel's region $\mathcal{F}_{\ell}$. 

\begin{figure} 
\centering\includegraphics[width=1\columnwidth]{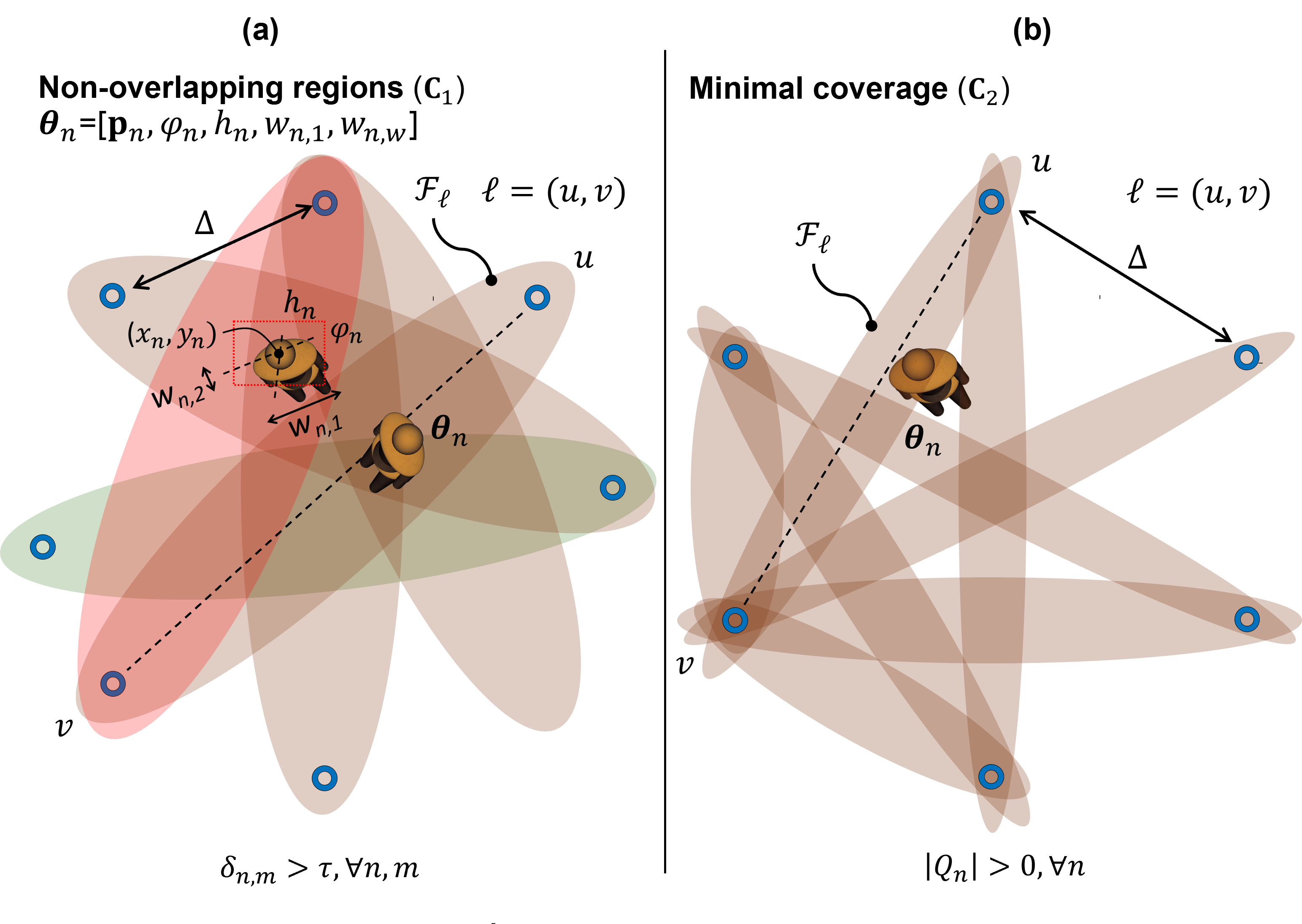} \caption{\label{c1c2} Figure (a) illustrates condition $C_{1}$, where the
resolvability of a target $n$ depends on the difference between the
sets of Fresnel's region $F_{\ell}$ that fully cover targets $n$ and
$m$. The Jaccard distance $\delta_{n,m}$ in (\ref{eq:jaccard})
indicates how unique the sets $Q_{n}$ and $Q_{m}$ (which represent the
Fresnel's region covering targets $n$ and $m$) are. The
blue and red Fresnel's regions highlight an example of zones that the
two targets do not share, allowing them to be uniquely identified as $\delta_{n,m}>\tau$.
Figure (b) instead depicts condition $C_{2}$, where the target lies
outside all Fresnel's regions, making it unresolved. This scenario corresponds
to $|Q_{n}|=0$, meaning no links fully cover target $n$.}
\end{figure}

\subsection{Conditions on target resolution}

Based on diffraction model considerations, a target $T_n$ located in
the monitored area $\mathcal{X}$ is considered resolvable
if it meets two main conditions, namely $\mathbf{C_{\mathrm{1}}}$ and $\mathbf{C_{\mathrm{2}}}$, which are related to the properties
of the corresponding Fresnel's regions. These conditions are visually represented in Fig.~\ref{c1c2}.
\begin{itemize}
\item \emph{Non-overlapping regions} ($\mathbf{C_{\mathrm{1}}}$). As shown in Fig.~\ref{c1c2}(a), condition
$\mathbf{C_{\mathrm{1}}}$ requires that the set $Q_{n}$ of links whose Fresnel's regions completely
cover target $n$ is unique. A similarity (or distance) metric is
used to measure such uniqueness: the metric quantifies the distance
of the set $Q_{n}$ from another $Q_{m}$, with $m\neq n$. Two targets
$m$ and $n$ are thus labeled as distinguishable only if the distance
metric of two corresponding link sets, $Q_{m}$ and $Q_{n}$, is larger
than an assigned tolerance value $\tau$. For two targets $n$ and $m$, we adopt the Jaccard distance~\cite{Jaccard} (here $|\cdot|$ is the cardinality of a set):
\begin{equation}
\delta_{n,m}=1-\frac{|Q_{n}\cap Q_{m}|}{|Q_{n}\cup Q_{m}|},\label{eq:jaccard}
\end{equation}
which quantifies the similarity between $Q_{n}$ and $Q_{m}$. For
a given threshold $\tau\geq0$, the condition $\delta_{n,m}>\tau$
corresponds to targets $n$ and $m$ being resolvable, as they possess
enough links to ensure the uniqueness of sets $Q_{n}$ and $Q_{m}$.
The resolvability condition can be evaluated as:
\begin{equation}
\Theta_{1}(n)=\prod_{\substack{m=1\\
m\neq n
}
}^{N}1_{\delta_{n,m}>\tau},\label{eq:c1}
\end{equation}
with $\Theta_{1}(n)=1$ indicating that the condition $\mathbf{C_{\mathrm{1}}}$
is verified and $0$ otherwise.

\item \emph{Minimal coverage} ($\mathbf{C}_{2}$). 
As shown in Fig.~\ref{c1c2}(b), the second condition
\emph{$\mathbf{C}_{2}$} verifies whether the target is covered by
at least one link. For a target $T_n$ to be considered resolvable,
there must be at least one link for which the target position $\mathbf{p}_{n}$
is inside its Fresnel's region. This condition is thus expressed as
$\sum_{\ell\in\mathcal{L}}\mathbf{1}_{\mathbf{p}_{n}\in\mathcal{F}_{\ell}}>0$,
which is equivalent to check if the set $Q_{n}$ is non-empty or
\begin{equation}
\Theta_{2}(n)=\mathbf{1}_{|Q_{n}|>0},\label{eq:c2b}
\end{equation}
with $\Theta_{2}(n)=1$ indicating that the condition $\mathbf{C}_{2}$
is verified and $0$ otherwise.
 
\end{itemize}

\subsection{Number of resolvable targets}

A target $T_n$ is considered resolvable if both \textbf{$\mathbf{C_{\mathrm{1}}}$}
and \textbf{$\mathbf{C_{\mathrm{2}}}$} conditions are met. Considering
the above, we can express the maximum number $\hat{N}\leq N$ of resolvable
targets as:


\begin{equation}
\hat{N}=\sum_{n=1}^{N}\underset{\mathbf{C_{\mathrm{1}}}\,(\ref{eq:c1})}{\underbrace{\Theta_{1}(n)}}\hspace{-0.5pt}\cdot\hspace{-0.5pt}\underset{\mathbf{C_{\mathrm{2}}}\,(\ref{eq:c2b})}{\underbrace{\Theta_{2}(n)}}\hspace{-0.5pt}+\hspace{-0.5pt}\sum_{n=1}^{N}\underset{\textrm{Correction factor}}{\underbrace{\frac{\Theta_{2}(n)}{\mathlarger{\sum}_{\substack{m=1\\
m\neq n
}
}^{N}\mathbf{1}_{\delta_{n,m}\leq\tau}}}}.\label{eq:limit}
\end{equation}

\noindent The first part of (\ref{eq:limit}) applies (\ref{eq:c1}) and (\ref{eq:c2b}) that correspond to the conditions \textbf{$\mathbf{C_{\mathrm{1}}}$}
and \textbf{$\mathbf{C_{\mathrm{2}}}$}, respectively. The second part introduces
a correction factor to adjust for cases where several targets overlap
in the same Fresnel's region. Further details are given in the Appendix. 
 The proposed limit takes into account
both the configuration of the environment and the overlapping between
the Fresnel's regions according to the specific antenna and link deployment
as well as the carrier frequency. 

Notice that (\ref{eq:limit}) allows to evaluate the average counting accuracy via classical Monte Carlo approach:

\begin{equation}
\frac{1}{n_{\text{s}}} \sum_{i=1}^{n_{\text{s}}} \mathbf{1}_{\hat{N}^{(i)} = N},
\label{eq:accuracy}
\end{equation}
where $n_{\text{s}}$ is the number of trials and $\hat{N}^{(i)}$ is the number of resolvable targets over the $i$-th trial, with $i=1,\ldots,n_{\text{s}}$. Each trial represents a random set of subject locations uniformly distributed within the area  $\mathcal{X}$.

In the examples of Sect.~\ref{sec:Numerical-results-and}, we analyze the impact of the tolerance threshold level $\tau$ on the resolvable targets and compare the bounds with a ML approach to people counting discussed in the following section.

\section{Graph Neural Network architecture\label{sec:Graph-Neural-Network}}

\begin{figure*}
\centering \includegraphics[scale=0.52]{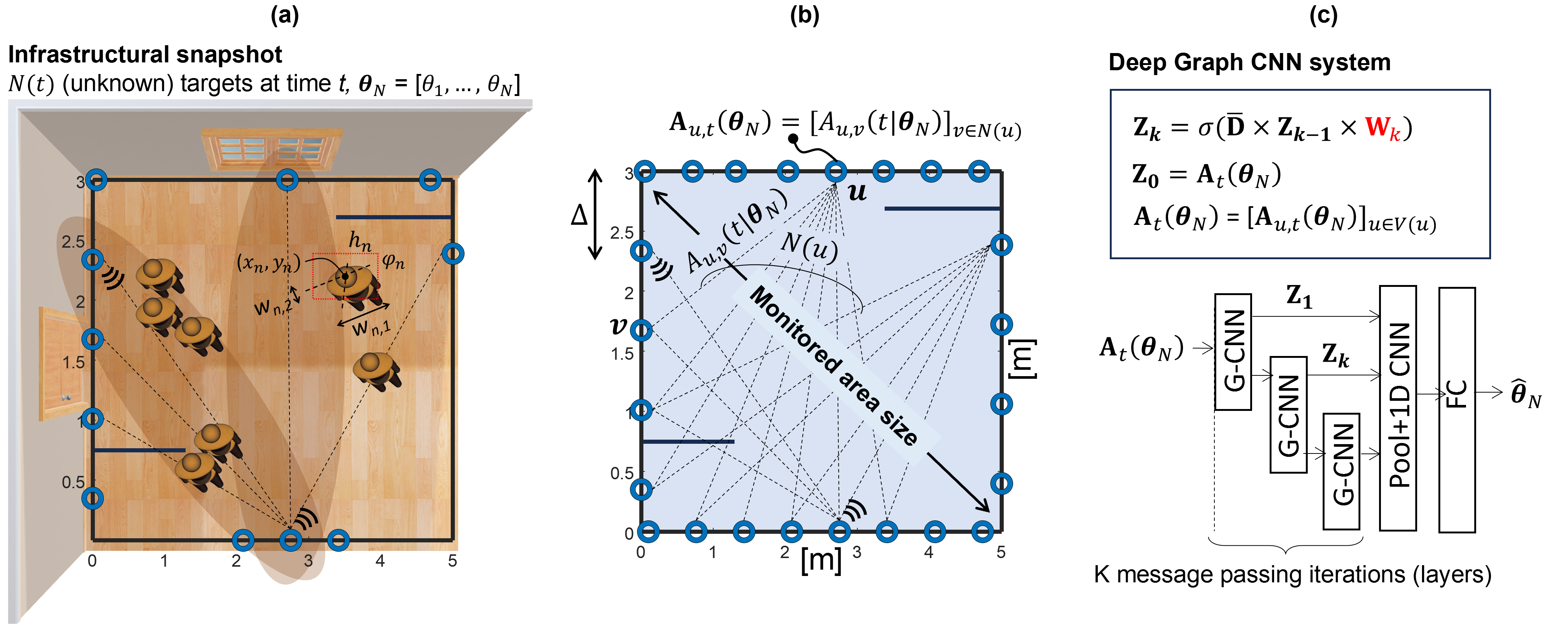} \protect\caption{\label{intro} Infrastructure snapshot scenario: (a) multi-target body
counting through RF signals inspection; (b) EM simulation environment with main parameters; (c) network structure and DGCNN model.}
\end{figure*}

A Deep Graph Convolutional Neural Network (DGCNN) model is adopted
in the following sections for motion classification. The DGCNN implements a message
passing scheme in which the measured attenuations $\mathbf{A}_{u,t}$,
observed at each node $u$ at time $t$, are exchanged between neighbor 
nodes according to the graph $\mathcal{G}$ and updated using neural
networks. The collection and processing of attenuations $\mathbf{A}_{u,t}$ is visually represented in Figures \ref{intro}(a) and \ref{intro}(b), respectively. Using the input graph $\mathcal{G}=(\mathcal{V},\mathcal{L})$,
along with a set of features $\mathbf{A}_{u,t}$,
the node $u$ embeddings $\mathbf{z}_{u}^{(k)}$, at iteration $k>0$,
are obtained iteratively as: 
\begin{equation}
\begin{cases}
\begin{aligned}
\mathbf{z}_{u}^{(0)}={} & \mathbf{A}_{u,t}\\
\mathbf{z}_{u}^{(k)}={} & \sigma(\mathbf{W}_{k}^{(0)}\mathbf{z}_{u}^{(k-1)}+\mathbf{W}_{k}^{(1)}\sum_{v\in\mathcal{N}(u)}\mathbf{z}_{v}^{(k-1)}+\mathbf{n}^{(k)}),
\end{aligned}
\end{cases}\label{eq:message_pass}
\end{equation}
where $\mathbf{W}_{k}=[\mathbf{W}_{k}^{(0)}$, $\mathbf{W}_{0}^{(1)}]$
are learnable matrices, $\sigma(\cdot)$ denotes an element wise non-linearity,
in our case the $\tanh(\cdot)$ function, while $\mathbf{n}^{(k)}$ is the trainable
bias term. At each iteration $k$ of the GNN, node $u$ takes as input
the set of embeddings $\mathbf{z}_{v}^{(k-1)}$ of the 1-hop neighborhood
$v\in\mathcal{N}(u)$ and combines them with the local embeddings
$\mathbf{z}_{u}^{(k-1)}$ to generate the updated ones. Notice that,
for all the nodes, the initial embeddings at $k=0$ are set to $\mathbf{z}_{u}^{(0)}=\mathbf{A}_{u,t}$
as in (\ref{eq:features}). On each iteration, the nodes are thus
set to mutually exchange, aggregate and update their local EM field
observations.

The DGCNN iterations $k=1,...,K$ correspond to the layers of classical convolutional
neural networks (CNN), therefore the parameters $\mathbf{W}_{k}$
can be trained separately for each layer/iteration or shared. As depicted in Fig. \ref{intro}(c), $K$ iterations of the GNN message passing are implemented first. Then, the output
of the final layer is used to collect the embeddings for each node: $\mathbf{Z}_{K}=\left[ \mathbf{z}_{u}^{(K)}\right] _{u\in\mathcal{V}}.$
In compact matrix form and using (\ref{eq:message_pass}), it is 
\begin{equation}
\mathbf{Z}_{K}=\sigma(\overline{\mathbf{D}}\times\mathbf{Z}_{K-1}\times\mathbf{W}_{K}+\mathbf{N}_{K}),\label{eq:compact}
\end{equation}
with $\mathbf{Z}_{0}=\mathbf{A}_{t}(\mathbf{\theta}_{N})$ and $\overline{\mathbf{D}}=\mathbf{D}+\mathbf{I}$
where $\mathbf{I}$ is the identity matrix, while $\mathbf{D}$ is the adjacency matrix defined as $\mathbf{D}\left[u,v\right]=1$,
$\mathrm{iff}$ $(u,v)\in\mathcal{L}$, $0$ elsewhere. Notice that after $k=K$ iterations
every node embedding contains information about its $k$-hop neighborhood.
The number $K$ of iterations is critical in real-time RF sensing,
while we also expect the degree distribution and the number of deployed
nodes to affect this choice.

\begin{table}[tp]
\protect\caption{\label{dgcnn}DGCNN implementation and main parameters}

\begin{centering}
\begin{tabular}{|l|l|}
\hline 
\textbf{Layer}  & \textbf{Configuration} \tabularnewline
 & \tabularnewline
\hline 
\multirow{4}{*}{\textbf{$K=4$ iterations} }  & $32$ units \tabularnewline
 & 32 units \tabularnewline
 & $32$ units \tabularnewline
 & $1$ unit \tabularnewline
\hline 
Conv1D  & $16$ filters, kernel size $96$ \tabularnewline
\hline 
Conv1D  & $32$ filters, kernel size $5$ \tabularnewline
\hline 
FC\#1  & 128 units \tabularnewline
\hline 
FC\#2  & $0\leq\widehat{N}$ $\leq20$ \tabularnewline
\hline 
\end{tabular}
\par\end{centering}
\medskip{}
\end{table}

\subsection{Implementation considerations}\label{implementation}

In the following, we highlight some relevant considerations about the
DGCNN network implementation. The graph network structure is based on the
StellarGraph library available in~\cite{stellar}. The software used for training the DGCNN model and defining the simulation environment is based on Python (version $\geq3.8$) and the TensorFlow package (version $\geq2.5$). As depicted in Tab.~\ref{dgcnn}, $K=4$ iterations are 
selected for the DGCNN model, with the adjacency normalization taken
from~\cite{aaai}-\cite{iclr}. The graph convolutional layers have
units shown in the Tab.~\ref{dgcnn}, while $\mathrm{tanh(\cdot)}$ activations are used as $\sigma(\cdot)$.
After GNN, three neural network layers are deployed, namely a 1D convolutional
layer (Conv1D) with $16$ filters, followed by a max pooling layer. Then, a second
Conv1D layer with $32$ filters is followed by two fully connected
(FC) layers with the last one used for regression, namely
to estimate the number of targets co-present $1\leq\widehat{N}\leq N=20$. The model
footprint is about $1.2$~MB while the observed inference time ($30$~ms)
is measured on a typical AI-based industrial IoT device (Jetson Nano IoT device
model) equipped with a 12-core ARM CPU and a low-power NVIDIA Maxwell GPU. 

The inputs of the DGCNN model correspond to the graph represented adjacency $\mathbf{D}$
and the node features matrices $\mathbf{A}_{u,t}$. According to the \emph{infrastructural snapshot} application case, the DGCNN model can be trained with attenuation samples $\mathbf{A}_{u,t}$ obtained either from simulations based on the MAM or the C-MAM models (see Sect.~\ref{sec:Numerical-results-and}), or from measurements (see Sect.~\ref{casestudy}). We consider a square network topology
highlighted in Fig.~\ref{intro}(a) and Fig. \ref{intro}(b): the network nodes are arranged along
the perimeter of a rectangular monitored area $\mathcal{X}$ in a
linear topology, where both the node spacing $\triangle$ and the size
$A$ of the area $\mathcal{X}$ can be varied, as well as the corresponding set of
active radio links $\mathcal{L}$, which define the adjacency matrix $\mathbf{D}$. As depicted in Fig.~\ref{intro}(c), the attenuation samples $\mathbf{A}_{u,t}$ are then fed into the node embeddings $\mathbf{Z}_{0}=\mathbf{A}_{t}=\left\{ \mathbf{A}_{u,t}\right\} _{u\in\mathcal{V}}$ at iteration $k=0$, followed by $K$ message passing steps.

\par\addvspace{2ex}

\subsection{Network deployment and scenarios\label{subsec:scenarios}}

In the following validation and tests we assume a variable area size ranging from $5$m$\times$$5$m (area $A=25$~sqm) up to $10$m$\times$$10$m (area $A=100$~sqm). A varying number of antennas is deployed around the perimeter of the monitored area $\mathcal{X}$, with spacing between adjacent antennas defined as $\triangle$. The attenuation samples $\mathbf{A}_{u,t}$ are obtained by EM simulations (Sect.~\ref{sec:em_model}) and assuming two carrier frequencies, namely $2.4$~GHz ($\lambda=12.5$~cm) and
$5.8$~GHz ($\lambda=5$~cm) representative of different Industrial, Scientific, and Medical (ISM) bands. Three different body models corresponding to Subject A, Subject B and Subject C are also considered as characterized by different physical dimensions (height $h_{n}$, anteroposterior and lateral dimensions $w_{n,1}$, $w_{n,2}$), namely: 
\begin{itemize}
	\item Subject A: $h_{n}=2$~m, $w_{n,1}=0.65$~m, $w_{n,2}=0.25$~m, 
	\item Subject B: $h_{n}=1.6$~m, $w_{n,1}=0.55$~m, $w_{n,2}=0.25$~m, 
	\item Subject C: $h_{n}=1.4$~m, $w_{n,1}=0.55$~m, $w_{n,2}=0.25$~m. 
\end{itemize}
These dimensions are designed to replicate those of a human body with different characteristics and are then evaluated in the experimental analysis of the following sections. 

\begin{figure}
\centering\includegraphics[scale=0.5]{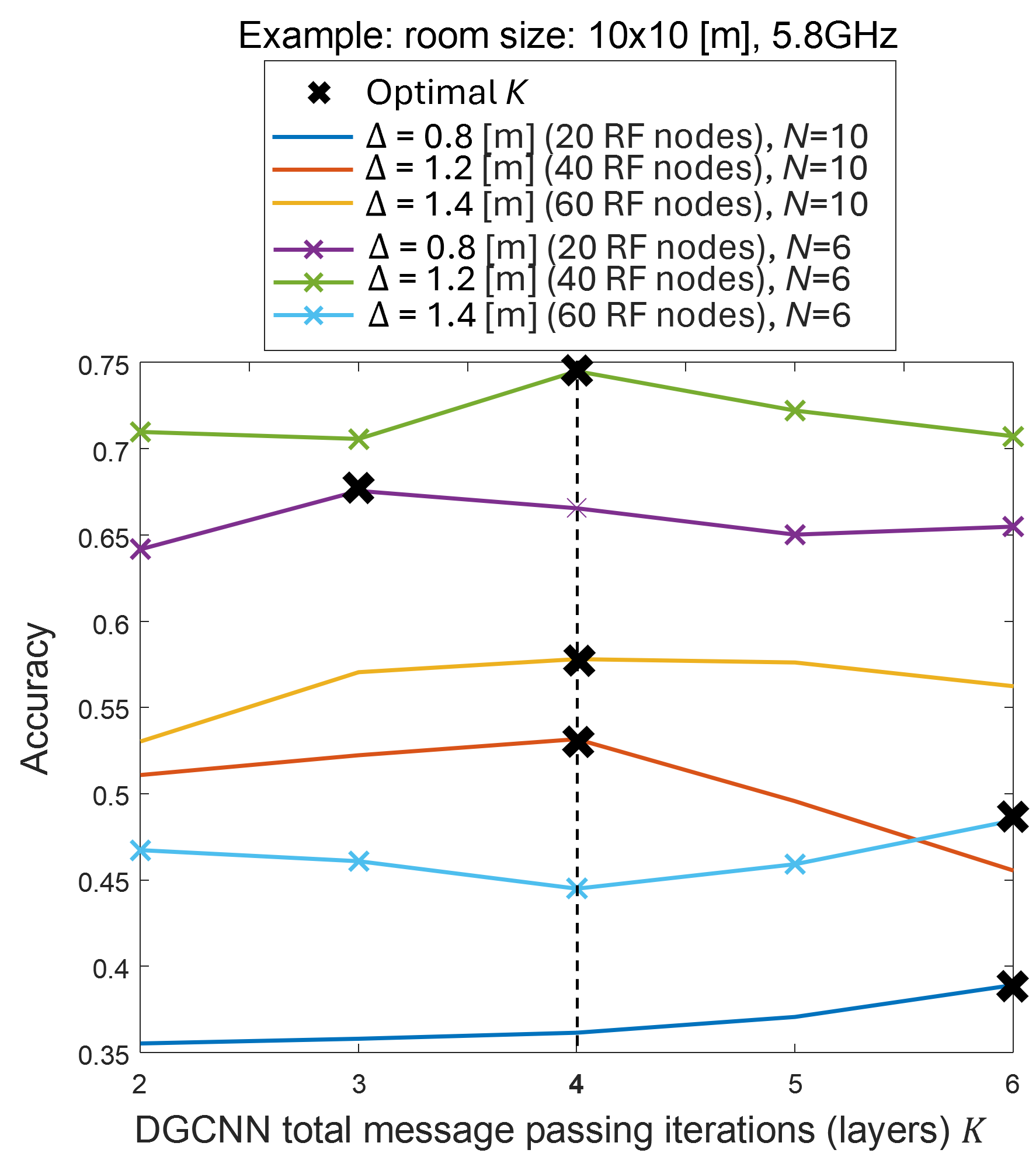} \caption{\label{iterations} Impact of the number of iterations $K$ of the
DGCNN network on the counting accuracy in two example scenarios consisting
of $N=10$ and $N=6$ targets in a $A=100$ sqm area space. Three networks
are considered with varying RF node and antenna spacing $\triangle=[0.8$~m,
$1.2$~m, $1.4$~m{]}, corresponding to $\left|\mathcal{V}\right|=20$, $\left|\mathcal{V}\right|=40$ and $\left|\mathcal{V}\right|=60$ RF nodes.
The optimal number of iterations is reported for each case (cross markers). The
DGCNN network is trained using the simulated attenuation values $\mathbf{A}_{u,t}$ corresponding to $750\times N$
random subject positions in the area $A$, for each proposed configuration.}
\end{figure}

Fig.~\ref{iterations} shows the effect of the number of iterations $K$ of the DGCNN network
on the accuracy performance in two selected scenarios consisting of $N=6$ and $N=10$ subjects moving randomly in a $A=100$~sqm area space. We consider three networks with varying number of RF nodes regularly deployed over the perimeter of the rectangular monitored area and different spacings $\triangle=[0.8$~m,
$1.2$~m, $1.4$~m{]}. The figure demonstrates that setting the number of iterations to $K=4$ strikes a reasonable balance between reducing real-time computation costs and achieving a minimum accuracy ($>0.5$) when the number of RF nodes is sufficient. For the following tests, we will set the number of iterations to $4$.

\begin{figure*}
\centering\includegraphics[width=\textwidth]{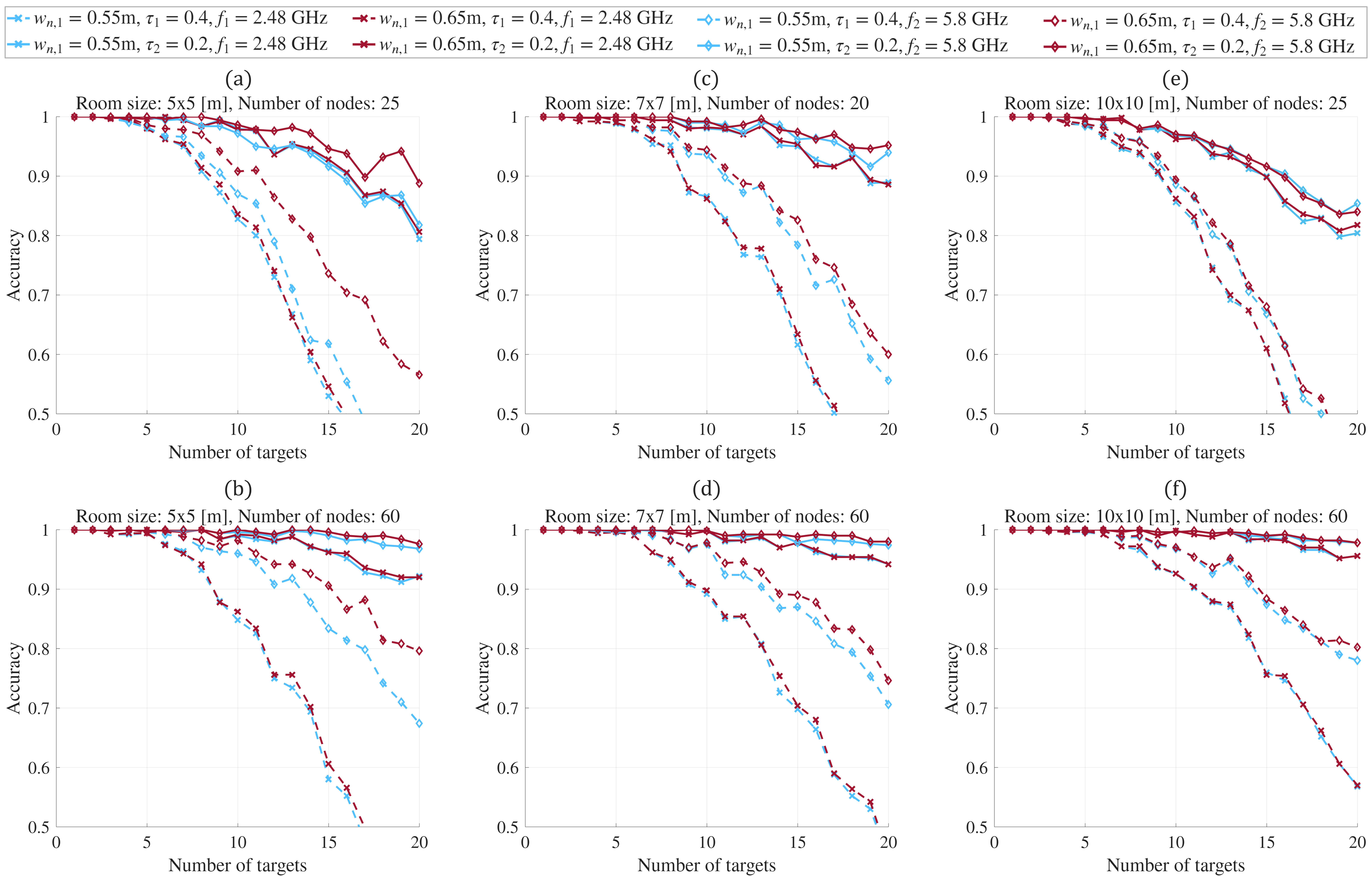} \caption{\label{fig4} Accuracy in estimating the true number of targets as a function of the actual number of targets, for different configurations and room sizes (5~m $\times$ 5~m, 7~m $ \times$ 7~m, and 10~m $\times$ 10~m). The results are obtained from the theoretical bounds proposed in Sect.~\ref{sec:limits} and are shown for different features: Jaccard distance threshold selected as $\tau_1 = 0.4$ (dashed line) and $\tau_2 = 0.2$ (solid line), operating frequency $f_1 = 2.48$~GHz (cross marker) and $f_2=5.8$~GHz (diamond marker) and target width $w_{n,1} = 0.55$~m (blue) and $w_{n,1} = 0.65$~m (red). For the $5$~m$\times5$~m and $10$~m$\times10$~m rooms, simulations are performed with $25$ and $60$ nodes; for the 7~m $\times$ 7~m room, $20$ and $60$ nodes are deployed.}
\end{figure*}

\section{Results from simulated environments\label{sec:Numerical-results-and}}

The section discusses and analyzes the proposed MAM and C-MAM diffraction
models and the theoretical bounds proposed in Sect.~\ref{sec:limits}. The goal is to assess their practical applicability in determining subject discrimination performance, namely the maximum number of subjects that can be reliably resolved via RF sensing, across various configurations. 

As discussed in the following, the counting accuracy
is obtained for variable area sizes and varying number of antennas (nodes), or graph vertexes $\left|\mathcal{V}\right|$. 
In all the cases considered
below, we evaluate the possibility of discriminating from $1$ to
$N$ targets, where $N$ is configurable up to a maximum of $N=20$, considered reasonable for crowd sensing applications~\cite{cianca}. 

\subsection{Analysis of accuracy bounds in dense graphs}

In the following, we analyze the accuracy bounds (\ref{eq:limit}) proposed in Sect.~\ref{sec:limits} and their validity under dense graph structures. To compute the accuracy (\ref{eq:accuracy}), the number $n_{\text{s}}$ of trials is set to $n_{\text{s}}=500$. Fig.~\ref{fig4} reports the accuracy bounds as a function of the number of targets, namely from $N=1$ to $N=20$, across different configurations of room size, antenna node density, target size ($w_{n,1}$, $w_{n,2}$), different threshold values ($\tau$ = 0.2, 0.4) of the Jaccard distance as defined in (\ref{eq:jaccard}), and both ISM band operating frequencies ($f_1=2.48$~GHz, $f_2=5.8$~GHz, i.e., WiFi). The six subplots of Fig.~\ref{fig4} correspond to:

\begin{enumerate}[label=(\alph*)]
\item Room size $5$m$\times5$m, $\left|\mathcal{V}\right|=25$ nodes
\item Room size $5$m$\times5$m, $\left|\mathcal{V}\right|=60$ nodes
\item Room size $7$m$\times 7$m, $\left|\mathcal{V}\right|=20$ nodes
\item Room size $7$m$\times7$m, $\left|\mathcal{V}\right|=60$ nodes
\item Room size $10$m$\times 10$m, $\left|\mathcal{V}\right|=25$ nodes
\item Room size $10$m$\times 10$m, $\left|\mathcal{V}\right|=60$ nodes
\end{enumerate}

The observed trends in Fig.~\ref{fig4} are a direct consequence of the theoretical resolvability bounds derived in Sect.~\ref{sec:limits}, namely in \eqref{eq:limit} and \eqref{eq:accuracy}. By imposing a minimum accuracy threshold of $90\%$, some general insights can be drawn. In the smallest room configuration ($5$m$\times$$5$m), subplot (b) shows that a denser deployment of $60$ nodes allows the system to maintain high accuracy up to $N=10$ targets with dimension $w_{n,1} = 0.65$~m, consistently with the fact that denser graphs provide more independent RSS measurements and thus higher resolvability. Conversely, when target dimensions become smaller or targets overlap, the Fresnel regions associated with different links increasingly intersect, reducing the number of unique link target associations and thus degrading performance.

For the medium-size room ($7$m$\times 7$m), subplot (d) shows that the same number of nodes ($\left|\mathcal{V}\right|=60$) supports accurate estimation of up to $N=12-14$ targets, with large subject dimensions and minimal overlap. However, reducing node density to $\left|\mathcal{V}\right|=20$, as in subplot (c), causes a sharp drop in accuracy: fewer nodes reduce the graph connectivity, hence limiting the number of resolvable targets.

Finally, in the largest room configuration ($10$m$\times10$m), subplot (f) demonstrates that up to $N=14-15$ targets can be reliably discriminated when operating at higher frequency ($f_2 = 5.8$~GHz), with a dense deployment ($60$ nodes) and adequate target separation ($\tau=0.4$). The advantage of $f_2$ follows from its smaller wavelength, which results in narrower Fresnel regions thus improving resolvability. Accuracy further increases with $\tau=0.2$, since a lower Jaccard distance threshold models a scenario with less overlap among targets and therefore higher discriminability. On the other hand, when only $25$ nodes are available, as in subplot (e), accuracy drops below the $90\%$ threshold already at $N=8-10$, in agreement with the theoretical prediction that lower node density reduces resolvability.

Overall, from Fig.~\ref{fig4}, three consistent trends emerge, each supported by the proposed bounds:
\begin{enumerate}
\item A larger target/subject width (i.e., anteroposterior dimension $w_{n,1} = 0.65$~m) improves classification accuracy, since larger bodies intersect more Fresnel's regions and create more distinguishable link signatures.
\item The Jaccard distance threshold $\tau$ tunes the degree of overlap among targets: a smaller $\tau$ (i.e., $0.2$) leads to improved discrimination, in line with the bounds that predict higher resolvability when unique link sets exist.
\item Operating at higher frequency ($f_2 = 5.8$~GHz) increases accuracy, as the narrower Fresnel's regions reduce spatial ambiguity and enable finer discrimination at higher $N$.
\end{enumerate}

Moreover, comparing subplots (b), (d), and (f) with (a), (c), and (e), we observe that larger environments can in principle support up to $N=20$ targets with accuracy $\geq90\%$, but only under the conditions predicted by the bounds: namely, dense deployments (>$60$ nodes to cover $A=100$ sqm) and operation at higher frequency band ($f_{2}$).
These considerations highlight the practical role of the bounds for pre-deployment network design: they provide a framework to calculate the node density, select the operating frequency, and the number of resolvable targets based on a desired accuracy.

\begin{table*}[tp]
\protect\caption{\label{modelA}Accuracy analysis for varying area size ranging from
$A=100$ sqm to $A=25$ sqm, and with RSS in the $2.4$~GHz and $5.8$~GHz bands, respectively.
Subjects A, B and C are characterized by different physical characteristics
as detailed in Sect.~\ref{subsec:scenarios}. Accuracy is obtained for varying number of
antennas with corresponding spacing $\triangle$ in {[}m{]}. Results
are obtained with MAM and C-MAM diffraction models discussed in Sect.~\ref{sec:em_model}.}

\begin{centering}
\begin{tabular}{c|c|c|c|c|c|c|c|}
\cline{3-8}
\multicolumn{1}{c}{} &  & \multicolumn{2}{c|}{\textbf{Subject A}} & \multicolumn{2}{c|}{\textbf{Subject B}} & \multicolumn{2}{c|}{\textbf{Subject C}}\tabularnewline
\cline{3-8}
\multicolumn{1}{c}{} &  & \begin{cellvarwidth}[t]

\hspace{-0.18 cm}$\triangle=0.67$ m

\hspace{-0.13 cm}$(60$ nodes$)$
\end{cellvarwidth} & \begin{cellvarwidth}[t]

\hspace{-0.18 cm}$\triangle=1.60$ m

\hspace{-0.13 cm}$(25$ nodes$)$
\end{cellvarwidth} & \begin{cellvarwidth}[t]
\centering
\hspace{-0.18 cm}$\triangle=0.67$ m

\hspace{-0.13 cm}$(60$ nodes$)$
\end{cellvarwidth} & \begin{cellvarwidth}[t]
\centering
\hspace{-0.18 cm}$\triangle=1.60$ m

\hspace{-0.13 cm}$(25$ nodes$)$
\end{cellvarwidth} & \begin{cellvarwidth}[t]
\centering
\hspace{-0.18 cm}$\triangle=0.67$ m

\hspace{-0.13 cm}$(60$ nodes$)$
\end{cellvarwidth} & \begin{cellvarwidth}[t]
\centering
\hspace{-0.18 cm}$\triangle=1.60$ m

\hspace{-0.13 cm}$(25$ nodes$)$
\end{cellvarwidth}\tabularnewline
\multicolumn{1}{c}{} &  &  MAM | C-MAM & MAM | C-MAM & MAM | C-MAM & MAM | C-MAM & MAM | C-MAM & MAM | C-MAM\tabularnewline
\hline \hline
\multirow{5}{*}{\begin{cellvarwidth}[t]
\centering
$A=100$ sqm

$5.8$~GHz
\end{cellvarwidth}} & $N=3$ &\hspace{-0.37 cm} $0.97$ | $0.79$ &\hspace{-0.37 cm} $0.92$ | $0.58$ &\hspace{-0.37 cm} $0.98$ | $0.82$ &\hspace{-0.37 cm} $0.89$ | $0.59$ &\hspace{-0.37 cm} $0.95$ | $0.79$ &\hspace{-0.37 cm} $0.87$ | $0.47$\tabularnewline
\cline{2-8}
 & $N=6$ &\hspace{-0.37 cm} $0.90$ | $0.68$ &\hspace{-0.37 cm} $0.78$ | $0.37$ &\hspace{-0.37 cm} $0.90$ | $0.61$ &\hspace{-0.37 cm} $0.77$ | $0.40$ &\hspace{-0.37 cm} $0.87$ | $0.66$ &\hspace{-0.37 cm} $0.75$ | $0.39$\tabularnewline
\cline{2-8}
 & $N=10$ &\hspace{-0.37 cm} $0.77$ | $0.49$ &\hspace{-0.37 cm} $0.66$ | $0.32$ &\hspace{-0.37 cm} $0.82$ | $0.51$ &\hspace{-0.37 cm} $0.64$ | $0.30$ &\hspace{-0.37 cm} $0.80$ | $0.52$ &\hspace{-0.37 cm} $0.63$ | $0.31$\tabularnewline
\cline{2-8}
 & $N=14$ &\hspace{-0.37 cm} $0.72$ | $0.44$ &\hspace{-0.37 cm} $0.57$ | $0.26$ &\hspace{-0.37 cm} $0.71$ | $0.43$ &\hspace{-0.37 cm} $0.54$ | $0.25$ &\hspace{-0.37 cm} $0.69$ | $0.43$ &\hspace{-0.37 cm} $0.52$ | $0.27$\tabularnewline
\cline{2-8}
 & $N=20$ &\hspace{-0.37 cm} $0.67$ | $0.35$ &\hspace{-0.37 cm} $0.52$ | $0.22$ &\hspace{-0.37 cm} $0.66$ | $0.33$ &\hspace{-0.37 cm} $0.50$ | $0.20$ &\hspace{-0.37 cm} $0.64$ | $0.37$ &\hspace{-0.37 cm} $0.52$ | $0.21$\tabularnewline
\hline \hline
\multirow{5}{*} {\begin{cellvarwidth}[t]
\centering
$A=100$ sqm

$2.4$~GHz
\end{cellvarwidth}} & $N=3$ &\hspace{-0.37 cm} $0.97$ | $0.87$ &\hspace{-0.37 cm} $0.91$ | $0.69$ &\hspace{-0.37 cm} $0.99$ | $0.87$ &\hspace{-0.37 cm} $0.94$ | $0.61$ &\hspace{-0.37 cm} $0.99$ | $0.87$ &\hspace{-0.37 cm} $0.88$ | $0.66$\tabularnewline
\cline{2-8}
 & $N=6$ &\hspace{-0.37 cm} $0.92$ | $0.68$ &\hspace{-0.37 cm} $0.78$ | $0.49$ &\hspace{-0.37 cm} $0.95$ | $0.70$ &\hspace{-0.37 cm} $0.77$ | $0.48$ &\hspace{-0.37 cm} $0.90$ | $0.66$ &\hspace{-0.37 cm} $0.79$ | $0.48$\tabularnewline
\cline{2-8}
 & $N=10$ &\hspace{-0.37 cm} $0.82$ | $0.55$ &\hspace{-0.37 cm} $0.65$ | $0.37$ &\hspace{-0.37cm} $0.86$ | $0.55$ &\hspace{-0.37 cm} $0.69$ | $0.37$ &\hspace{-0.37 cm} $0.83$ | $0.53$ &\hspace{-0.37 cm} $0.65$ | $0.35$\tabularnewline
\cline{2-8}
 & $N=14$ &\hspace{-0.37 cm} $0.73$ | $0.49$ &\hspace{-0.37 cm} $0.56$ | $0.33$ &\hspace{-0.37 cm} $0.77$ | $0.47$ &\hspace{-0.37 cm} $0.58$ | $0.32$ &\hspace{-0.37 cm} $0.75$ | $0.43$ &\hspace{-0.37 cm} $0.58$ | $0.30$\tabularnewline
\cline{2-8}
 & $N=20$ &\hspace{-0.37 cm} $0.69$ | $0.41$ &\hspace{-0.37 cm} $0.55$ | $0.27$ &\hspace{-0.37 cm} $0.69$ | $0.37$ &\hspace{-0.37 cm} $0.55$ | $0.27$ &\hspace{-0.37 cm} $0.72$ | $0.37$ &\hspace{-0.37 cm} $0.52$ | $0.26$\tabularnewline
\hline \hline
\multicolumn{2}{c|}{} & \begin{cellvarwidth}[t]
\centering
\hspace{-0.18 cm}$\triangle=0.47$ m

\hspace{-0.13 cm}$(60$ nodes$)$
\end{cellvarwidth} & \begin{cellvarwidth}[t]
\centering
\hspace{-0.18 cm}$\triangle=1.4$ m

\hspace{-0.13 cm}$(20$ nodes$)$
\end{cellvarwidth} & \begin{cellvarwidth}[t]
\centering
\hspace{-0.18 cm}$\triangle=0.47$ m

\hspace{-0.13 cm}$(60$ nodes$)$
\end{cellvarwidth} & \begin{cellvarwidth}[t]
\centering
\hspace{-0.18 cm}$\triangle=1.4$ m

\hspace{-0.13 cm}$(20$ nodes$)$
\end{cellvarwidth} & \begin{cellvarwidth}[t]
\centering
\hspace{-0.18 cm}$\triangle=0.47$ m

\hspace{-0.13 cm}$(60$ nodes$)$
\end{cellvarwidth} & \begin{cellvarwidth}[t]
\centering
\hspace{-0.18 cm}$\triangle=1.4$ m

\hspace{-0.13 cm}$(20$ nodes$)$
\end{cellvarwidth}\tabularnewline
\multicolumn{1}{c}{} &  & MAM | C-MAM & MAM | C-MAM & MAM | C-MAM & MAM | C-MAM & MAM | C-MAM & MAM | C-MAM\tabularnewline
\hline \hline
\multirow{5}{*}{\begin{cellvarwidth}[t]
\centering
$A=49$ sqm

$5.8$~GHz
\end{cellvarwidth}} & $N=3$ &\hspace{-0.37cm} $0.97$ | $0.85$ &\hspace{-0.37cm} $0.89$ | $0.57$ &\hspace{-0.37cm} $0.98$ | $0.81$ &\hspace{-0.37cm} $0.89$ | $0.56$ &\hspace{-0.37cm} $0.98$ | $0.82$ &\hspace{-0.37cm} $0.90$ | $0.57$\tabularnewline
\cline{2-8}
 & $N=6$ &\hspace{-0.37cm} $0.88$ | $0.66$ &\hspace{-0.37cm} $0.75$ | $0.39$ &\hspace{-0.37cm} $0.90$ | $0.64$ &\hspace{-0.37cm} $0.74$ | $0.38$ &\hspace{-0.37cm} $0.90$ | $0.64$  &\hspace{-0.37cm} $0.76$ | $0.39$\tabularnewline
\cline{2-8}
 & $N=10$ &\hspace{-0.37cm} $0.70$ | $0.54$ &\hspace{-0.37cm} $0.61$ | $0.30$ &\hspace{-0.37cm} $0.80$ | $0.51$ &\hspace{-0.37cm} $0.61$ | $0.30$ &\hspace{-0.37cm} $0.79$ | $0.53$ &\hspace{-0.37cm} $0.62$ | $0.29$\tabularnewline
\cline{2-8}
 & $N=14$ &\hspace{-0.37cm} $0.66$ | $0.45$ &\hspace{-0.37cm} $0.53$ | $0.26$ &\hspace{-0.37cm} $0.59$ | $0.45$ &\hspace{-0.37cm} $0.53$ | $0.24$ &\hspace{-0.37cm} $0.69$ | $0.44$ &\hspace{-0.37cm} $0.53$ | $0.24$\tabularnewline
\cline{2-8}
 & $N=20$ &\hspace{-0.37cm} $0.59$ | $0.37$ &\hspace{-0.37cm} $0.49$ | $0.21$ &\hspace{-0.37cm} $0.58$ | $0.36$ &\hspace{-0.37cm} $0.50$ | $0.20$ &\hspace{-0.37cm} $0.58$ | $0.36$ &\hspace{-0.37cm} $0.50$ | $0.22$\tabularnewline
\hline \hline
\multirow{5}{*}{\begin{cellvarwidth}[t]
\centering
$A=49$ sqm

$2.4$~GHz
\end{cellvarwidth}} & $N=3$ &\hspace{-0.37cm} $0.98$ | $0.88$ &\hspace{-0.37cm} $0.89$ | $0.65$ &\hspace{-0.37cm} $0.99$ | $0.86$ &\hspace{-0.37cm} $0.89$ | $0.67$ &\hspace{-0.37cm} $0.98$ | $0.86$ &\hspace{-0.37cm} $0.89$ | $0.64$\tabularnewline
\cline{2-8}
 & $N=6$ &\hspace{-0.37cm} $0.92$ | $0.69$ &\hspace{-0.37cm} $0.78$ | $0.46$ &\hspace{-0.37cm} $0.89$ | $0.75$ &\hspace{-0.37cm} $0.74$ | $0.49$ &\hspace{-0.37cm} $0.92$ | $0.70$ &\hspace{-0.37cm} $0.78$ | $0.46$\tabularnewline
\cline{2-8}
 & $N=10$ &\hspace{-0.37cm} $0.77$ | $0.58$ &\hspace{-0.37cm} $0.65$ | $0.37$ &\hspace{-0.37cm} $0.82$ | $0.62$ &\hspace{-0.37cm} $0.60$ | $0.39$ &\hspace{-0.37cm} $0.82$ | $0.55$ &\hspace{-0.37cm} $0.65$ | $0.36$\tabularnewline
\cline{2-8}
 & $N=14$ &\hspace{-0.37cm} $0.71$ | $0.48$ &\hspace{-0.37cm} $0.57$ | $0.32$ &\hspace{-0.37cm} $0.70$ | $0.51$ &\hspace{-0.37cm} $0.54$ | $0.32$ &\hspace{-0.37cm} $0.73$ | $0.45$ &\hspace{-0.37cm} $0.59$ | $0.31$\tabularnewline
\cline{2-8}
 & $N=20$ &\hspace{-0.37cm} $0.65$ | $0.42$ &\hspace{-0.37cm} $0.52$ | $0.28$ &\hspace{-0.37cm} $0.65$ | $0.43$ &\hspace{-0.37cm} $0.50$ | $0.27$ &\hspace{-0.37cm} $0.66$ | $0.39$ &\hspace{-0.37cm} $0.53$ | $0.26$\tabularnewline
\hline \hline
\multicolumn{1}{c}{} &  & \begin{cellvarwidth}[t]
\centering
\hspace{-0.18cm}$\triangle=0.33$ m

\hspace{-0.13cm}$(60$ nodes$)$
\end{cellvarwidth} & \begin{cellvarwidth}[t]
\centering
\hspace{-0.18cm}$\triangle=1$ m

\hspace{-0.13cm}$(25$ nodes$)$
\end{cellvarwidth} & \begin{cellvarwidth}[t]
\centering
\hspace{-0.18cm}$\triangle=0.33$ m

\hspace{-0.13cm}$(60$ nodes$)$
\end{cellvarwidth} & \begin{cellvarwidth}[t]
\centering
\hspace{-0.18cm}$\triangle=1$ m

\hspace{-0.13cm}$(25$ nodes$)$
\end{cellvarwidth} & \begin{cellvarwidth}[t]
\centering
\hspace{-0.18cm}$\triangle=0.33$ m

\hspace{-0.13cm}$(60$ nodes$)$
\end{cellvarwidth} & \begin{cellvarwidth}[t]
\centering
\hspace{-0.18cm}$\triangle=1$ m

\hspace{-0.13cm}$(25$ nodes$)$
\end{cellvarwidth}\tabularnewline
\multicolumn{1}{c}{} &  & MAM | C-MAM & MAM | C-MAM & MAM | C-MAM & MAM | C-MAM & MAM | C-MAM & MAM | C-MAM\tabularnewline
\hline \hline
\multirow{5}{*}{\begin{cellvarwidth}[t]
\centering
$A=25$ sqm

$5.8$~GHz
\end{cellvarwidth}} & $N=3$ &\hspace{-0.37cm} $0.95$ | $0.86$ &\hspace{-0.37cm} $0.94$ | $0.64$ &\hspace{-0.37cm} $0.98$ | $0.87$ &\hspace{-0.37cm} $0.93$ | $0.67$ &\hspace{-0.37cm} $0.96$ | $0.89$ &\hspace{-0.37cm} $0.93$ | $0.68$\tabularnewline
\cline{2-8}
 & $N=6$ &\hspace{-0.37cm} $0.83$ | $0.71$ &\hspace{-0.37cm} $0.81$ | $0.44$ &\hspace{-0.37cm} $0.85$ | $0.73$ &\hspace{-0.37cm} $0.81$ | $0.49$ &\hspace{-0.37cm} $0.82$ | $0.71$ &\hspace{-0.37cm} $0.78$ | $0.48$\tabularnewline
\cline{2-8}
 & $N=10$ &\hspace{-0.37cm} $0.69$ | $0.58$ &\hspace{-0.37cm} $0.66$ | $0.36$ &\hspace{-0.37cm} $0.66$ | $0.58$ &\hspace{-0.37cm} $0.66$ | $0.38$ &\hspace{-0.37cm} $0.72$ | $0.57$ &\hspace{-0.37cm} $0.69$ | $0.37$\tabularnewline
\cline{2-8}
 & $N=14$ &\hspace{-0.37cm} $0.57$ | $0.47$ &\hspace{-0.37cm} $0.59$ | $0.32$ &\hspace{-0.37cm} $0.53$ | $0.47$ &\hspace{-0.37cm} $0.55$ | $0.31$ &\hspace{-0.37cm} $0.61$ | $0.49$ &\hspace{-0.37cm} $0.60$ | $0.31$\tabularnewline
\cline{2-8}
 & $N=17$ &\hspace{-0.37cm} $0.53$ | $0.44$ &\hspace{-0.37cm} $0.58$ | $0.28$ &\hspace{-0.37cm} $0.51$ | $0.45$ &\hspace{-0.37cm} $0.56$ | $0.29$ &\hspace{-0.37cm} $0.62$ | $0.43$ &\hspace{-0.37cm} $0.57$ | $0.27$\tabularnewline
\hline \hline
\multirow{5}{*}{\begin{cellvarwidth}[t]
\centering
$A=25$ sqm

$2.4$~GHz
\end{cellvarwidth}} & $N=3$ &\hspace{-0.37cm} $0.96$ | $0.92$ &\hspace{-0.37cm} $0.94$ | $0.74$ &\hspace{-0.37cm} $0.97$ | $0.88$ &\hspace{-0.37cm} $0.94$ | $0.69$ &\hspace{-0.37cm} $0.98$ | $0.92$ &\hspace{-0.37cm} $0.94$ | $0.71$\tabularnewline
\cline{2-8}
 & $N=6$ &\hspace{-0.37cm} $0.88$ | $0.75$ &\hspace{-0.37cm} $0.85$ | $0.55$ &\hspace{-0.37cm} $0.89$ | $0.68$ &\hspace{-0.37cm} $0.82$ | $0.52$ &\hspace{-0.37cm} $0.87$ | $0.77$ &\hspace{-0.37cm} $0.82$ | $0.54$\tabularnewline
\cline{2-8}
 & $N=10$ &\hspace{-0.37cm} $0.75$ | $0.64$ &\hspace{-0.37cm} $0.73$ | $0.42$ &\hspace{-0.37cm} $0.76$ | $0.54$ &\hspace{-0.37cm} $0.72$ | $0.41$ &\hspace{-0.37cm} $0.71$ | $0.59$ &\hspace{-0.37cm} $0.73$ | $0.45$\tabularnewline
\cline{2-8}
 & $N=14$ &\hspace{-0.37cm} $0.64$ | $0.52$ &\hspace{-0.37cm} $0.61$ | $0.38$ &\hspace{-0.37cm} $0.65$ | $0.51$ &\hspace{-0.37cm} $0.64$ | $0.36$ &\hspace{-0.37cm} $0.69$ | $0.50$ &\hspace{-0.37cm} $0.61$ | $0.37$\tabularnewline
\cline{2-8}
 & $N=17$ &\hspace{-0.37cm} $0.59$ | $0.48$ &\hspace{-0.37cm} $0.57$ | $0.35$ &\hspace{-0.37cm} $0.65$ | $0.44$ &\hspace{-0.37cm} $0.57$ | $0.33$ &\hspace{-0.37cm} $0.62$ | $0.47$ &\hspace{-0.37cm} $0.59$ | $0.34$\tabularnewline
\hline \hline
\end{tabular}
\par\end{centering}
\medskip{}
 
\end{table*}

\subsection{DGCNN model analysis with generated samples}

\begin{figure*}
\centering\includegraphics[width=\textwidth]{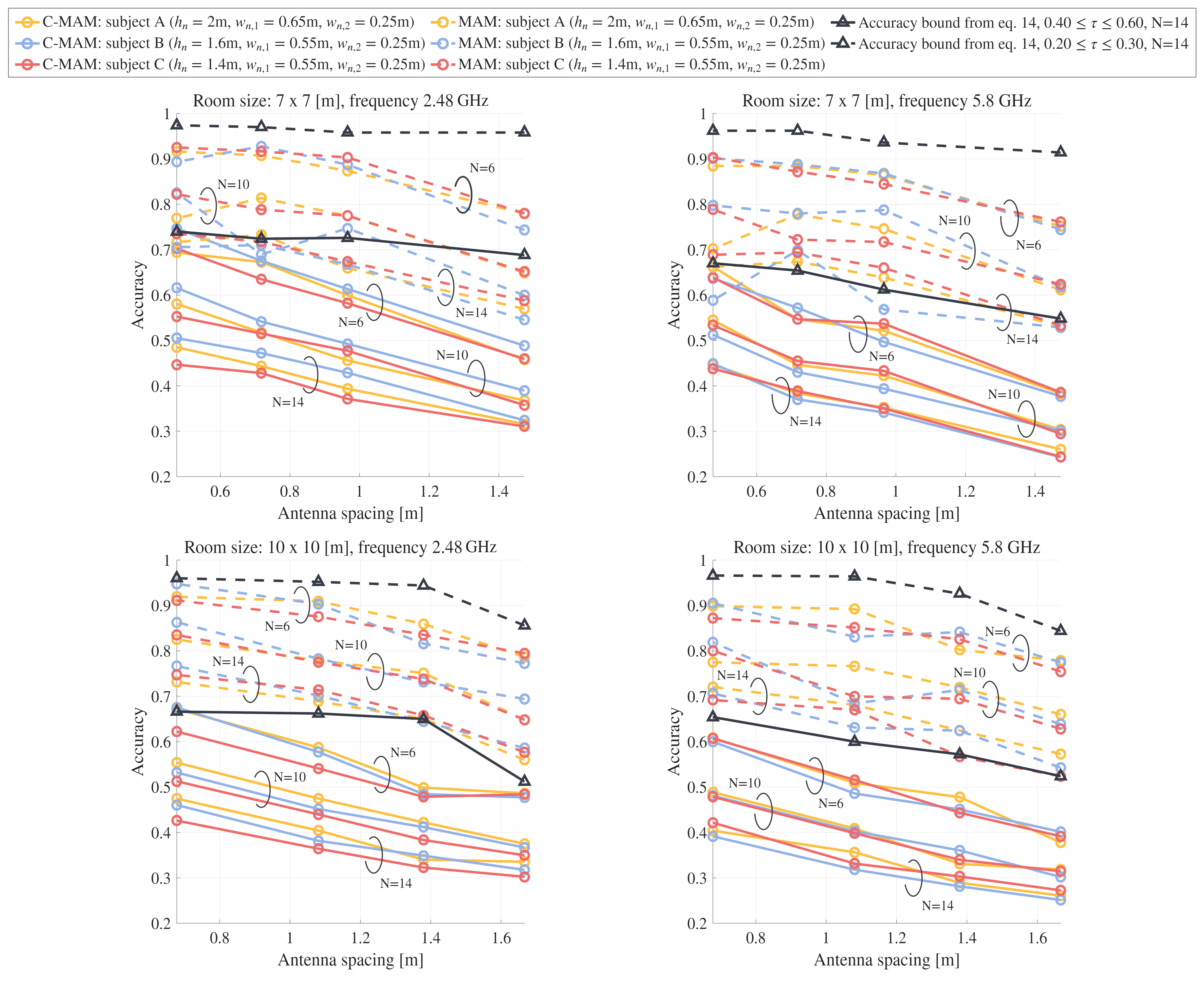} \caption{\label{models_comparison} Accuracy analysis with DGCNN model and generated RF attenuation samples from MAM and C-MAM models. Prediction of the number $N$ of subjects moving in a 10 m~$\times$~10~m (bottom) and 7 m~$\times$~7~m (top) area. Subjects A, B and C are considered and they are characterized by different physical dimensions respectively represented in yellow, blue and red. Accuracy is evaluated as a function of antenna spacing $\Delta$ at $f_1=2.48$~GHz (left) and $f_2=5.8$~GHz (right). Comparison of MAM and C-MAM accuracy is shown for $N = 6,\,10,\,14$ subjects. Solid lines with circular markers indicate C-MAM results, dashed lines with circular markers indicate MAM results, while grey lines with triangular markers represent the theoretical accuracy bounds from (\ref{eq:accuracy}) for $N=14$ targets. The bounds are shown as solid or dashed depending on the chosen value of $\tau$ \eqref{eq:c1}, tuned to represent a meaningful bound in each case.}
\end{figure*}

\begin{figure*}
\centering \includegraphics[width=\textwidth]{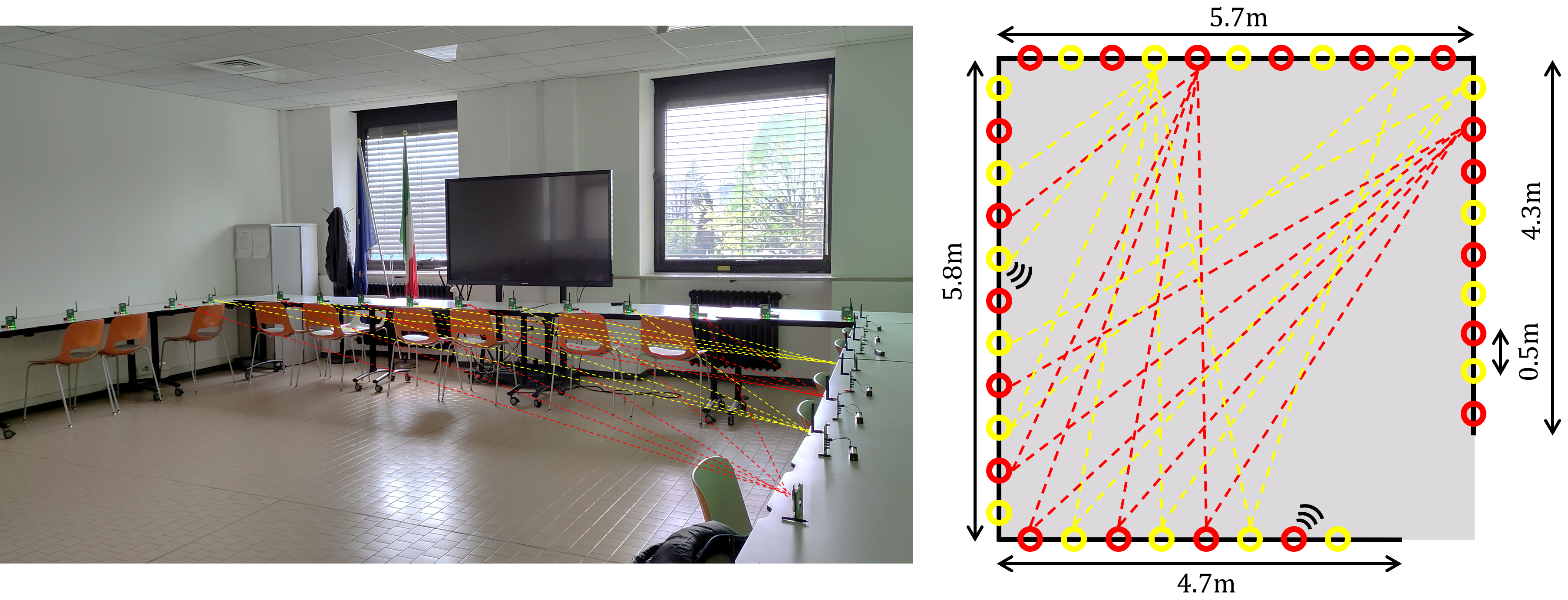} \protect\caption{\label{tests} Case study scenario: multi-target body counting area (left), network layouts (Network \#1 and \#2) with main geometric parameters (right). It is worth noticing that the combined network (\#1+\#2) is not fully connected.}
\end{figure*}

We now evaluate the system performance using the DGCNN model presented
earlier, trained on RF attenuation samples $A_{u,v}$ generated by the MAM and C-MAM models.
The goal is to assess the degree to which the proposed DGCNN processing system produces results consistent with the purely
geometric bounds presented in the previous section. 

Fig.~\ref{models_comparison} shows the accuracy in predicting the
number of targets co-present within a $A=100$~sqm (bottom) and
$A=49$~sqm (top) area. Three subjects (Subject A, Subject B, and Subject C) with different
physical dimensions, represented in yellow, blue and red, respectively, are considered. Accuracy is evaluated as a function of antenna antenna spacing
$\triangle$ [m] at two operating frequencies, namely $f_1=2.48$~GHz on the left and $f_2=5.8$~GHz on the right. The DGCNN graph model is trained using attenuation samples generated by the MAM (dashed lines with circular markers) and C-MAM (solid
lines with circular markers) models. Performance is compared for three different target densities: $N = 6,\,10\,,14$. Additionally, the figure includes the theoretical accuracy bounds derived from (\ref{eq:accuracy}) for $N=14$ (grey lines with triangular markers), shown as solid or dashed depending on the chosen value of the Jaccard distance threshold $\tau$ in \eqref{eq:c1}, tuned to represent a meaningful accuracy limit in each case. The selected range of $\tau$ values for Fig.~\ref{models_comparison} is $0.4 \leq \tau \leq 0.6$ for C-MAM bounds, and $0.2 \leq \tau \leq 0.3$ for MAM bounds. The accuracy results for all tested scenario are also reported in Tab.~\ref{modelA} which covers a wider range of conditions including target counts from $N=3$ up to $N=20$, multiple frequencies and different room sizes. 

By analyzing these results, some remarks are discussed here:
\begin{enumerate}
\item The antenna spacing $\triangle$ significantly impacts performance;
the accuracy generally improves as spacing decreases, however a
saturation effect is observed for the MAM model
when $\triangle<1$ meter.
\item The C-MAM model accounts for practical phenomena such as subject-induced ambiguity in RF sensing, resulting in more conservative (i.e., lower) accuracy estimates. This makes it a useful tool for identifying potential performance bottlenecks in the system design phase.
\item The theoretical bounds from (\ref{eq:limit}) (black lines with triangle markers) can be tuned via the Jaccard distance threshold $\tau$ to approximate both MAM and C-MAM behavior; 
\item Subjects with smaller body dimensions (e.g., Subject C) are more difficult to detect because it becomes harder to associate them with multiple distinct Fresnel's ellipses (and hence, RF links). As a consequence, ambiguity in the counting process is more likely to occur. This limitation is particularly evident in the C-MAM results and is also supported by the theoretical accuracy bounds. 
\end{enumerate}


\begin{table}[tp]
\protect\caption{\label{network}Deployed networks parameters for experiments.}

\begin{centering}
\begin{tabular}{l|l|l|l|}
\cline{2-4} 
 & \textbf{Network \#1} & \textbf{Network \#2}  & \textbf{Networks (\#1+\#2)}\tabularnewline
\hline 
$\triangle$ {[}m{]} & $1.0$ & $1.0$ & $0.5$\tabularnewline
\hline 
$L$ & $171$ & $190$ & $342$\tabularnewline
\hline 
$\left|\mathcal{V}\right|$ & $19$ & $20$ & $38$\tabularnewline
\hline 
$f_{c} {[\text{GHz}]}$ & $2.41$ (ch. $12$) & $2.43$ (ch. $16$) & $-$\tabularnewline
\hline 
\end{tabular}
\par\end{centering}
\medskip{}
 
\end{table}





\section{Indoor case study and validation\label{casestudy}}

In this section, we present the results of an indoor case study performed in the $2.4$~GHz band. The study aims to validate the established bounds analyzed previously and the possibility to adopt the proposed MAM and C-MAM models to predict the performance during the pre-deployment phase.

Fig.~\ref{tests} illustrates an experimental deployment of the network nodes. The network architecture employs IoT devices operating in the $2.4$~GHz frequency band and adhering to the IEEE 802.15.4
physical layer specification, a standard commonly adopted in industrial
deployments~\cite{davoli}. Note that the validation assumes static RF links (aside from body-induced variations), which may limit applicability in highly dynamic environments. In the following, we first describe the implemented network for synchronous, time-slotted collection of attenuation samples. Next, we describe the experimental validation and the comparative analysis with the proposed accuracy bounds and models.

\subsection{RF sensing network implementation and data collection}

The network operates using a synchronous, time-slotted approach. The RF nodes are programmed to transmit Physical Protocol Data Unit frames, namely beacons, at pre-assigned time slots with duration $0.5$~ms and guard time of $0.15$~ms. During each time slot, the network nodes measure the RSS of the received beacons. Collected samples are then transmitted to a central Access Point (AP), which acts as the hub for data processing and analysis. A snapshot, namely one RSS measurement for each link of the graph $\mathcal{G}$, is collected by the AP every $60$~ms, which corresponds to one IEEE 802.15.4 superframe\footnote{Notice that the transmitter (TX) and receiver (RX) roles can be switched in each snapshot according to the assigned graph topology.}. The protocol is implemented on a testbed using nodes equipped with the low-power SoC JN5189 transceiver from NXP~\cite{datasheet}. 
Although the Media Access Control (MAC) sublayer of the IEEE 802.15.4 standard has been customized for RF sensing, it still maintains full compatibility with the base standard, and several popular machine-type communication protocols such as RFC 4944 (6LoWPAN), Time Slotted Channel Hopping (TSCH) solutions, i.e., 6TiSCH~\cite{154}.

Narrowband operation is assumed, as is commonly encountered in machine-type communication systems where subcarrier-level amplitude or RSS information can be used. The system operating frequency can be selected at run time, before deployment using standard IEEE 802.15.4 channels. Multiple networks might therefore operate in parallel over different, i.e., adjacent channels, without suffering from interference. Two networks (red and yellow nodes in Fig.~\ref{tests})
with characteristics detailed in the Tab.~\ref{network} were deployed in the monitored area shown in Fig.~\ref{tests} (left).
Both networks operate on two distinct frequencies, namely $2.41$~GHz (channel 12 of the IEEE 802.15.4 standard) and $2.43$~GHz (channel 16). The combined network integrates a total of $\left|\mathcal{V}\right|=38$
antennas (nodes/edges) with $\triangle=0.5$ m spacing and $L=342$
links. It has to be noted that the combined network can be approximated as a single network working at $2.42$~GHz (channel 14) since the attenuation effects due to the targets for a frequency difference of 10~MHz are negligible~\cite{rampa-2017}.

The network nodes are designed to collect the RSS measurements $P_{u,v}$, in dBm (decibel-milliwatts), defined as:
\begin{equation}
P_{u,v}=\left\{ \begin{array}{ll}
P_{u,v}(\emptyset)+w_{R} & \;\textrm{\ensuremath{N=0}}\\
P_{u,v}(\emptyset)-A_{u,v}+w_{T} & \;\textrm{\ensuremath{N>0} subjects \ensuremath{\mathbf{\boldsymbol{\theta}}_{N}} },
\end{array}\right.\label{eq:received_power}
\end{equation}
where $P_{u,v}(\emptyset)$ is the received power in the empty space ($N=0$) and
it is assumed to be known or measured in the reference scenario. The terms $w_{R}$ and and $w_{T}$ model the log-normal
multipath fading and the other disturbances~\cite{rampa-2022a,j-fieramosca-2023}.

Using the power measurements (\ref{eq:received_power}), the body induced attenuation for
link ($u,v$) is estimated as $\widehat{A}_{u,v}=P_{u,v}(\emptyset)-\mathrm{E_{\mathit{t}}}[P_{u,v}]$,
with $\mathrm{E_{\mathit{t}}}[\cdot]$ being the expected value over
consecutive snapshots. The attenuation terms $\widehat{\mathbf{A}}_{u,t}:=\left\{ \widehat{A}_{u,v}\right\} _{v\in\mathcal{N}(u)}$,
observed by each node $u$ at time $t$ and according to the graph $\mathcal{G}$, are collected by the AP and used as inputs for DGCNN training and inference.

\par\addvspace{2ex}

\subsection{Experimental validation}

In what follows, we compare the crowd sensing performance under two scenarios:
1) the DGCNN model $\mathbf{W}_{k}$, $k=1,...,K$ trained with \textit{measured}
attenuations $\widehat{\mathbf{A}}_{u,t}$ obtained from up to $N=7$
real subjects; 2) the DGCNN model now trained using attenuation samples
$\mathbf{A}_{u,t}$ \textit{generated} by C-MAM and MAM models that are set
to reproduce the environment in Fig.~\ref{tests} as well as the
subject physical dimensions. The objective is to evaluate the predictive
potential of the models in comparison to real-world measurements and
the specific setup. 

\begin{figure}
\centering\includegraphics[scale=0.34]{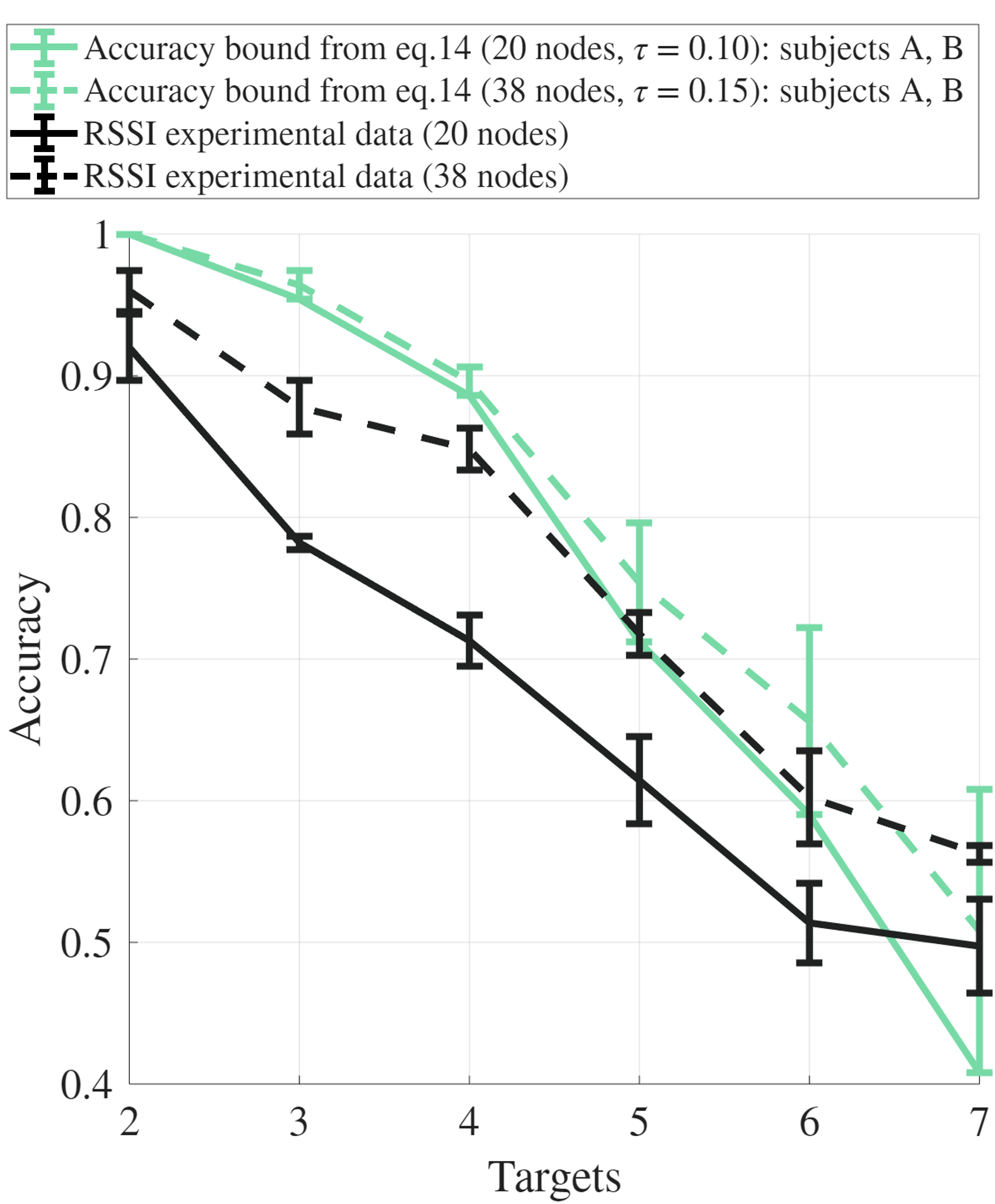} \caption{\label{experiment_results} Average counting accuracy of the DGCNN model for varying number of targets (from $2$ to $7$) using RSS experimental data (black lines with error bars), showing results for $20$ (solid line) and $38$ RF nodes (dashed line). Accuracy is averaged over $750$ snapshots/trials. For the same setups, we also reported the accuracy bounds derived from \eqref{eq:accuracy} (green lines), with the corresponding values of the Jaccard distance threshold $\tau$ \eqref{eq:c1} indicated in the legend. For the experimental data, error bars represent the variability over different trials; for the analytical bounds, they reflect the variability due to different physical characteristics of the subjects (Subject A, and B).}
\end{figure}

\begin{figure}
\centering\includegraphics[scale=0.33]{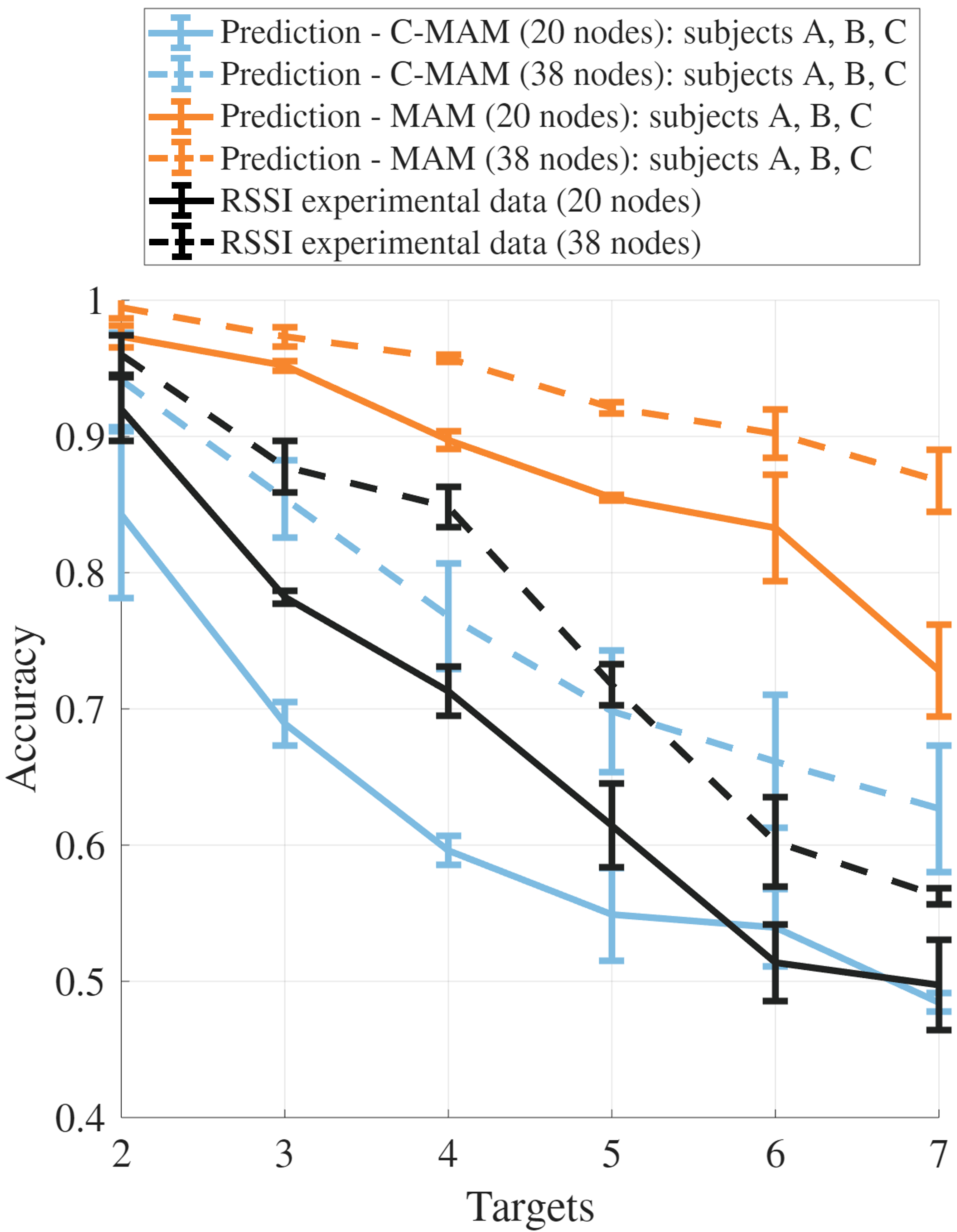} \caption{\label{experiment_results_b} Average counting accuracy of the DGCNN model for varying number of targets (from $2$ to $7$) and RF nodes ($20$ -- solid lines -- and $38$ -- dashed lines) using attenuation samples generated by C-MAM (blue) and MAM (orange) models. Predicted accuracy is compared with RSS experimental results (black curves) as reported in Fig.~\ref{experiment_results}. Error bars for C-MAM and MAM reflect the variability due to different subjects (Subject A, B, and C).}
\end{figure}

All the proposed scenario assume static RF links as in typical IoT smart-home environments. The DGCNN model is trained on multiple snapshots that capture the continuous, time-varying changes in the RF signals as the result of human movements. Fig.~\ref{experiment_results} reports the counting accuracy obtained with RSS experimental data (black lines with error bars), using $\left|\mathcal{V}\right|=20$ (solid line) and $\left|\mathcal{V}\right|=38$ (dashed line) nodes, together with the accuracy bounds derived from (\ref{eq:accuracy}) (green lines), whose Jaccard distance thresholds $\tau$ in \eqref{eq:c1} are indicated in the legend. For RSS measurements, the error bars represent the variability across training and validation sets obtained through 3-fold cross-validation.
Fig.~\ref{experiment_results_b} complements this analysis by comparing the same RSS results with the predictions of DGCNN models trained on the attenuation samples generated by the C-MAM (blue lines) and MAM (orange lines) models. For these predictions, the error bars reflect the variability due to different physical dimensions of the subjects, based on the body models Subject A, B, and C defined in Sect.~\ref{subsec:scenarios}.

From the joint examination of these two figures, the following observations can be drawn:

\begin{enumerate}

\item As depicted in Fig. ~\ref{experiment_results}, the accuracy bounds (\ref{eq:accuracy}) for $\left|\mathcal{V}\right|=20$ (solid lines) and $\left|\mathcal{V}\right|=38$ nodes (dashed lines) appear to be accurate for higher accuracy scenarios ($>0.5$). The slight discrepancy observed for $N=7$ targets may stem from the fact that the approximation \eqref{eq:limit} used to derive the accuracy bounds assumes identical targets, a condition that does not hold true for real-world RSS measurements.
\item The trend of the accuracy obtained through the RSS measurements is well aligned with the prediction of the C-MAM model as shown in Fig.~\ref{experiment_results_b},
while MAM model results can be considered as an upper bound to real performance.
\item For a limited number of deployed nodes/antennas ($\left|\mathcal{V}\right|=20$),
C-MAM predictions tend to slightly underestimate accuracy (by approximately
10$\%$) for target counts below $N=4$. 
This effect is likely due to the model inability to account for multipath effects, which might enhance performance in the specific environment.
\item When the number of nodes increases to $\left|\mathcal{V}\right|=38$, the predictions from C-MAM align more closely with the experimental data for intermediate values of $N$, but an overestimation of around $5\%$ can be observed for $N>5$.
\item The uncertainty related to the subjects' physical dimensions increases with the number of co-present targets, a trend that is consistently reflected in both model predictions and real measurements.
\end{enumerate}

\section{Conclusions\label{sec:Conclusions}}

The paper presented an EM-informed approach for estimating the number
of people in a monitored area using RF sensing in dense IoT network topologies.
We developed a composite multi-body additive EM model (C-MAM) that
accounts for relevant interactions of RF signals with multiple individuals,
and we derived practical limits on the number of resolvable targets
based on this model. The proposed resolvable conditions are tied to
the geometric properties of RF links and their corresponding Fresnel's
regions; they are designed to assess RF sensing performance,
specifically the average accuracy and the number of resolvable subjects, during
the pre-deployment stage. This assessment was carried out under
various propagation scenarios and considering different physical characteristics
of human bodies and crowd sizes. 

A deep graph convolutional neural network (DGCNN) system was trained
on synthetic data generated by the C-MAM model reproducing EM body
effects. As the number of subjects increases, body-induced EM
blockage affects more links, resulting in distinct graph populations.
The DGCNN architecture proved effective in predicting the number
of people while mitigating overfitting. The proposed bounds and limits
are also validated through experimental measurements of Received Signal
Strength (RSS), demonstrating the model's predictive potential during the pre-deployment or design stages.

Future work will focus on extending the generalization capability of the DGCNN model to arbitrary network deployments, where fewer RF nodes are available. In addition, the DGCNN architecture can be refined to incorporate temporal dynamics, namely time-series analysis of the RSS, which is crucial for achieving more accurate and robust real-time tracking of people movement. Finally, extending the C-MAM modeling approach to explicitly incorporate frequency-dependent effects i.e., for multi-band acquisitions, could enable frequency-diversity features and improve the robustness to multipath effects.

\section*{Appendix}

Considering a target $T_{n}$ and condition \textbf{$\mathbf{C}_{1}$,
}the product\textbf{{} $\prod_{m=1,m\neq n}^{N}\mathbf{1}_{\delta_{n,m}>\tau}$}{}
is $1$ when the target $T_n$ is distinguishable from the other subjects
$T_m$ (and $0$ otherwise): two targets can be thus separated when
the distance metric of two corresponding link sets $\delta_{n,m}>\tau$
is larger than an assigned tolerance value $\tau$. A minimal coverage,
as condition $\mathbf{C}_{2}$, must be also applied since the set
$Q_{n}$ should be non-empty $|Q_{n}|>0$. The product 
\begin{equation}
\Theta(n)=\underset{\Theta_{1}(n)}{\underbrace{\prod_{m=1,m\neq n}^{N}\mathbf{1}_{\delta_{n,m}>\tau}}}\cdot\underset{\Theta_{2}(n)}{\underbrace{\mathbf{1}_{|Q_{n}|>0}}}\label{eq:pr}
\end{equation}
is thus $1$ when the target $T_n$ is distinguishable according to
conditions $\mathbf{C}_{1}$ and $\mathbf{C}_{2}$ and $0$ otherwise.

Notice that a target for which the condition $\mathbf{C}_{2}$ is verified as $\Theta_{2}(n)=1$ but $\Theta_{1}(n)=0$, although not distinguishable
since $\Theta(n)=0$, still contribute to the total target count as
sharing the same region with other targets. We define 
\begin{equation}
\Psi(n)=\sum_{m=1,m\neq n}^{N}\mathbf{1}_{\delta_{n,m}\leq\tau}\label{eq:m}
\end{equation}
as the number of targets sharing the same region of space with subject
$T_n$, namely for which the distance metric $\delta_{n,m}\leq\tau$.
These targets are counted as one single blockage element and not resolvable.
The contribution of target $T_n$ to such total count is therefore $\frac{1}{\Psi(n)}$.
Using (\ref{eq:pr}) and (\ref{eq:m}), we can express the maximum
number $\hat{N}\leq N$ of resolvable targets as: 
\begin{equation}
\hat{N}=\sum_{n=1}^{N}\left[\Theta_{1}(n)\Theta_{2}(n)+\frac{\Theta_{2}(n)}{\Psi(n)}\right]\label{eq:limit-1}
\end{equation}
which corresponds to (\ref{eq:limit}).
\par\addvspace{1ex}

\vspace{-1cm}
\begin{IEEEbiographynophoto}{Federica Fieramosca }
	received the M.Sc. degree in Telecomunnication Engineering (cum laude)
	from Politecnico di Milano, Italy, in 2022. From November 2022, she
	is a PhD student in Information Technology at the Department of Electronics,
	Information and Bioengineering at Politecnico di Milano. She is currently
	working in the field of applied electromagnetics, focusing on electromagnetic
	propagation modelling for integrated sensing and communication.
\end{IEEEbiographynophoto}
\begin{IEEEbiographynophoto}{Vittorio Rampa }
	(Senior Member, IEEE) received the M.Sc degree (Hons.) in EE from
	Politecnico di Milano (POLIMI), Italy, in 1984. In 1986, he joined
	the Consiglio Nazionale delle Ricerche (CNR) as Researcher. From
	1999 to 2014, he has been Adjunct Professor at POLIMI where he has taught courses
	on software defined radio algorithms and architectures, and navigation
	systems. From 2001 to 2022, he has been Senior Researcher at the Institute
	of Electronics, Computer, and Telecommunication Engineering (CNR-IEIIT).
	In 2006, he co-founded the CNR-POLIMI spin-off Wisytech Srl where,
	till 2014, he served as CTO. Since 2018, he is member of the Steering
	Committee GTTS-4 of the National Cluster Fabbrica Intelligente. Since
	2023, he is Senior Research Associate at CNR-IEIIT. His research
	interests include signal processing algorithms and architectures for
	wireless communications; wireless sensor networks for IoT applications;
	radio vision methods for detection, localization, and activity recognition
	of people; body models and machine learning tools for radio passive sensing; federated learning algorithms for cooperative
	and green applications. 
\end{IEEEbiographynophoto}
\begin{IEEEbiographynophoto}{Michele D'Amico }
	(Senior Member, IEEE) Michele D'Amico graduated from Politecnico
	di Milano in 1990; in 1997 he received his "Ph.D. in Mathematics"
	from the University of Essex (UK). Since 2002 he has been Associate
	Professor in Applied Electromagnetics at DEIB (Politecnico di Milano).
	Research activities include antennas, electromagnetic wave propagation
	in the troposphere at frequencies above 10 GHz, and radarmeteorology.
	Since May 2013 he is responsible for the experimental activities of
	the Spino d'Adda laboratory of the Politecnico di Milano, dedicated
	to the experimental investigation of EM waves in the troposphere.
	He is Senior Member of the IEEE. He has several patents on antennas
	and has authored or co-authored more than 140 papers published in
	international journals or conference proceedings. 
\end{IEEEbiographynophoto}
\begin{IEEEbiographynophoto}{Stefano Savazzi }(Senior Member, IEEE) is a Senior Researcher at Consiglio Nazionale delle
	Ricerche (CNR), the Institute of Electronics, Computer and Telecommunication
	Engineering (IEIIT). He received the M.Sc. degree and the Ph.D. degree
	(Hons.) in ICT from the Politecnico di Milano, Italy, in 2004 and
	2008, respectively, and joined CNR in 2012. He was a Visiting Researcher
	with Uppsala University, in 2005 and University of California at San
	Diego in 2007. He has coauthored over 130 scientific publications
	(Scopus). His current research interests include distributed signal
	processing, distributed machine learning and networking aspects for the Internet
	of Things, radio localization and vision technologies. Dr. Savazzi
	was the recipient of the 2008 Dimitris N. Chorafas Foundation Award.
	He is principal investigator for CNR in Horizon EU projects Holden, TRUSTroke and the Doctoral Network SMARTTEST. He is also serving as Associate Editor for Frontiers
	in Communications and Networks, Wireless Communications and Mobile
	Computing and Sensors. 
\end{IEEEbiographynophoto}
\end{document}